\documentclass[%
 reprint,
 amsmath,amssymb,
 aps,
]{revtex4-1}

\usepackage{graphicx}
\usepackage{dcolumn}
\usepackage{bm}

\usepackage[utf8]{inputenc}
\usepackage[T1]{fontenc}
\usepackage{mathptmx}
\usepackage{etoolbox}
\usepackage[dvipsnames]{xcolor}
\usepackage{siunitx}
\usepackage{float}

\newcommand{\gammai}{\gamma_\text{i}}

\begin{document}

\preprint{APS/123-QED}

\title{Flexible Integration  of Gigahertz Nanomechanical Resonators with a Superconducting Microwave Resonator using a Bonded Flip-Chip Method}

\author{Sultan Malik}
\thanks{These authors contributed equally}
\author{Wentao Jiang}
\thanks{These authors contributed equally}
\author{Felix M. Mayor}
\thanks{These authors contributed equally}
\author{Takuma Makihara}
\author{Amir H. Safavi-Naeini}%
\email{safavi@stanford.edu}

\affiliation{Department of Applied Physics and Ginzton Laboratory, Stanford University, 348 Via Pueblo Mall, Stanford, California 94305, USA}

\date{\today}

\begin{abstract}
 We demonstrate strong coupling of gigahertz-frequency nanomechanical resonators to a frequency-tunable superconducting microwave resonator via a galvanically bonded flip-chip method. By tuning the microwave resonator with an external magnetic field, we observe a series of hybridized microwave-mechanical modes and report coupling strengths of $\sim \SI{15}{\mega\hertz}$ at cryogenic temperatures. The demonstrated multi-chip approach provides flexible rapid characterization and simplified fabrication, and could potentially enable coupling between a variety of quantum systems. Our work represents a step towards a plug-and-play architecture for building more complex hybrid quantum systems.
\end{abstract}

\maketitle

In the last decade, hybrid quantum systems have provided a playground for developing useful quantum technologies and exploring fundamental research in quantum mechanics, mesoscopic physics, and condensed-matter physics~\cite{Clerk2020}. In a hybrid quantum system, physical systems with distinct quantum degrees of freedom, such as microwave photons, optical photons, phonons, spins, and magnons, are combined to leverage their unique strengths~\cite{Clerk2020}. Of particular interest are the hybrid systems consisting of nanomechanical resonators which have potential applications in quantum-limited sensing~\cite{Mason2019}, storage of quantum information~\cite{Pechal2018, Hann2019}, and microwave-to-optical quantum transduction~\cite{jiang2022optically, Meesala2023, Weaver2022, mirhosseini2020superconducting, jiang2020efficient, han2020cavity} for building future superconducting qubits-based quantum networks~\cite{han2021microwave}. 

Achieving quantum control of a hybrid quantum system calls for quantum states to be manipulated faster than any decoherence rate. This translates to having large couplings between various modes or degrees of freedom in the system. In addition to large coupling, good frequency matching between the modes is required. This can be particularly challenging because of the fabrication complexity of a hybrid quantum system. Hence, an integration approach that allows some of the components to be fabricated and tested independently, and allows them to be assembled with ease and flexibility is particularly appealing. The field of circuit quantum electrodynamics (cQED) has explored two-dimensional (2D)~\cite{Blais2004, Wallraff2004}, three-dimensional (3D)~\cite{Paik2011, Wang2014}, and quasi-3D~\cite{satzinger2019simple, conner2021superconducting} integration approaches to couple qubits to readout resonators. Similarly, the field of circuit quantum acousto-dynamics (cQAD)~\cite{chu2017quantum, chu2018creation, arrangoiz2018coupling, satzinger2018quantum, arrangoiz2019resolving, sletten2019resolving, peterson2019ultrastrong, wollack2022quantum, bienfait2019phonon} has explored 3D~\cite{peterson2019ultrastrong} and quasi-3D integration approaches, like inductive coupling~\cite{satzinger2018quantum, bienfait2019phonon} and capacitive coupling~\cite{chu2017quantum, chu2018creation, wollack2022quantum} between a mechanical resonator and a qubit. In this work, we use a combination of 3D and quasi-3D approaches to couple gigahertz-frequency nanomechanical resonators to a superconducting microwave resonator via a  galvanically bonded flip-chip method. We read out the resonators using a 3D microwave cavity and report strong coupling at cryogenic temperatures. The demonstrated multi-chip architecture provides flexibility, simplified fabrication, and a convenient integration scheme for hybrid quantum systems.

\begin{figure}[tb]
\includegraphics[scale=1]{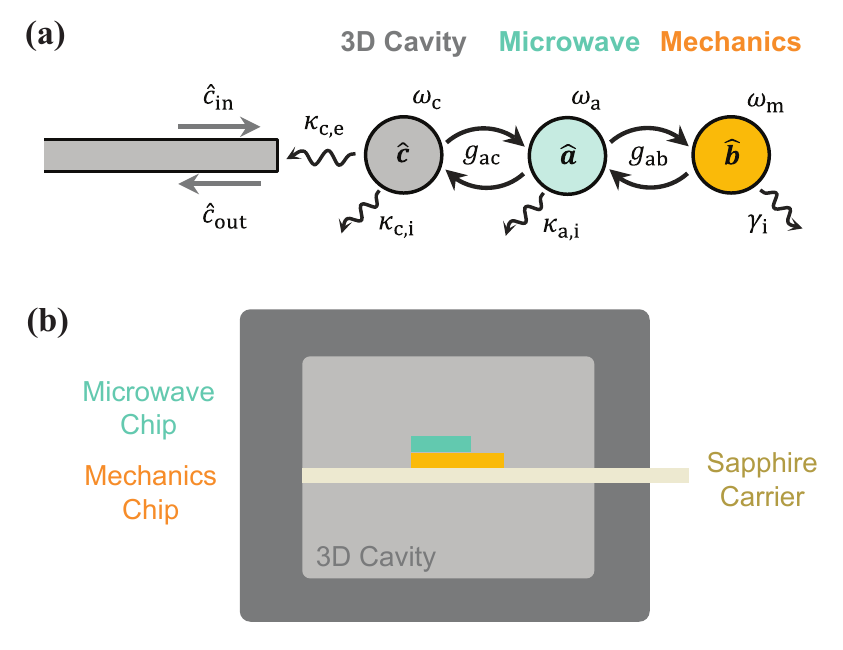}
\caption{\label{fig:overview} \textbf{Mode overview and schematic.} \textbf{(a)} Definition of modes and their coupling rates. Mechanical modes are coupled to a microwave mode that is coupled to a 3D cavity and read out via transmission measurement through the 3D cavity. \textbf{(b)} Schematic showing the physical placement of the chips containing the nanomechanical resonators and microwave resonator with respect to the 3D cavity. The mechanics chip and the microwave chip are galvanically bonded to each other. The pair is glued onto a sapphire carrier that is then placed inside the 3D cavity.}
\end{figure}

The system consists of a mechanical mode, a microwave mode of a chip-scale superconducting resonator, and the cavity mode of a machined copper cavity. The two physical interactions in this system are the linear piezoelectric interaction between the mechanical mode and the microwave mode, and the electromagnetic interaction between the on-chip microwave mode and the cavity mode (Fig.~\ref{fig:overview}(a)). The cavity is connected to a $\SI{50}{\ohm}$ microwave transmission line for measurement. The Hamiltonian describing the system is 
\begin{eqnarray} \label{eq1}
    \hat H / \hbar&=&\omega_\text{c} \hat{c}^{\dagger} \hat{c}+\omega_\text{a} \hat{a}^{\dagger} \hat{a}+\omega_\text{m} \hat{b}^{\dagger} \hat{b} \notag\\
    &&+g_\text{ab}\left(\hat{a}^{\dagger} \hat{b}+\hat{b}^{\dagger} \hat{a}\right)+g_\text{ac}\left(\hat{a}^{\dagger} \hat{c}+\hat{c}^{\dagger} \hat{a}\right),
\end{eqnarray}
where $\omega_{\text{c}}$, $\omega_{\text{a}}$, $\omega_{\text{m}}$ and $\hat{c}$, $\hat{a}$, $\hat{b}$ represent the frequencies and annihilation operators of the cavity mode, the microwave mode, and the mechanical mode, respectively. The coupling rate between the cavity and the microwave modes and the coupling rate between the microwave and the mechanical modes are represented by $g_\text{ac}$ and $g_\text{ab}$, respectively (Fig.~\ref{fig:overview}(a)). 

The three modes in the system represent resonators that are physically distinct objects and are all fabricated independently. The cavity mode arises in a macroscopic 3D piece of copper, which houses the chip-scale microwave and nanomechanical resonators fabricated on separate chips. The microwave chip and the mechanics chip are heterogeneously integrated in a flip-chip manner with galvanic contacts between them. The microwave-mechanics chip pair is then glued onto a sapphire carrier that is placed inside the 3D cavity (Fig.~\ref{fig:overview}(b)). The microwave resonator is electromagnetically coupled to the 3D cavity whereas the coupling between the microwave resonator and nanomechanical resonator is mediated via the galvanic contacts to electrodes that climb onto the nanomechanical resonator.

\begin{figure}[!htbp]
\includegraphics[scale=1]{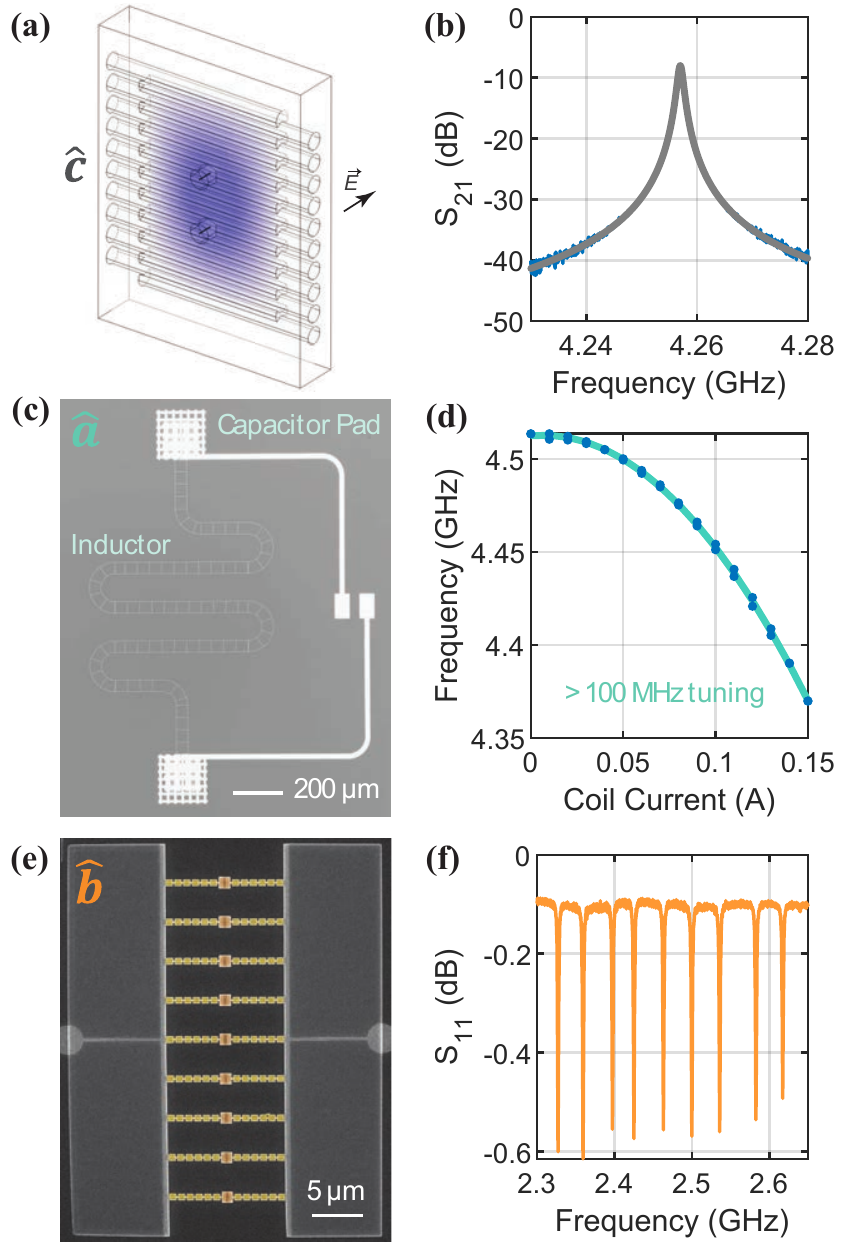}
\caption{\label{fig:modes}  \textbf{Physical implementation of the modes and their characterization.} \textbf{(a)} The 3D cavity is formed by drilling overlapping holes on two ends of an oxygen-free high thermal conductivity copper (OFHC) block resulting in a seamless 3D cavity with a hollow region which acts as the cavity mode volume. The shaded blue region in the FEA model corresponds to the magnitude of the simulated $\vec{E}$-field for the TE$_{101}$ mode of the cavity. \textbf{(b)} Transmission measurement of the 3D cavity at $T\sim \SI{4}{\kelvin}$ with a fundamental TE mode at $\omega_\text{c}/2\pi = \SI{4.257}{\giga\hertz}$. \textbf{(c)} Optical microscope image of the superconducting microwave resonator. The microwave resonator is composed of a pair of antenna-style capacitor pads and a ladder-style nanowire inductor with high kinetic inductance made of niobium titanium nitride (NbTiN). \textbf{(d)} The fundamental microwave resonance mode ($T\sim \SI{10}{\milli\kelvin}$) can be tuned ($> \SI{100}{\mega\hertz}$) by wirelessly tuning the kinetic inductance through an external magnetic field generated by current in a coil. The plot shows the overlay of both tuning up and tuning down curves. \textbf{(e)} False-color scanning electron microscope image of the suspended nanomechanical resonators. Each nanomechanical resonator consists of a defect site embedded in a phononic crystal that supports a complete phononic bandgap in the frequency range $\SI{2.3}-\SI{3.0}{\giga\hertz}$ on a highly piezoelectric thin-film lithium niobate (LN) platform. \textbf{(f)}. Reflection $S_{11}$ measurement of the nanomechanical resonators at room temperature.}
\end{figure}

The 3D microwave cavity, microwave resonator circuit, and mechanical resonators are shown in Fig.~\ref{fig:modes}. We fabricate the 3D cavity using the flute technique~\cite{chakram2021seamless}. A series of overlapping holes are drilled on two ends of a monolithic oxygen-free high thermal conductivity (OFHC) copper resulting in a seamless 3D cavity. The hollow region resulting from the overlap of the drilled holes acts as the majority of the cavity mode volume. In addition to the ease of assembly and higher quality factors due to the absence of seam loss, a design like this allows  various  other systems such as optical photons and atoms~\cite{Suleymanzade2020, Haroche2020} to be incorporated in the cavity, making it an appropriate design choice for a hybrid quantum system. Fig.~\ref{fig:modes}(a) shows a finite-element analysis (FEA) model showing the $\vec{E}$-field magnitude for the TE$_{101}$ mode of the cavity. The 3D cavity in this setup has a TE$_{101}$ mode at $\omega_\text{c}/2\pi = \SI{4.257}{\giga\hertz}$ as measured by the transmission measurement through the cavity at $T\sim \SI{4}{\kelvin}$ (Fig.~\ref{fig:modes}(b)). To make the 3D cavity reusable and adaptable to various microwave-mechanics chip pairs, it is desirable to tune its fundamental mode frequency. We utilize a combination of sapphire rods (for coarse control) and sapphire strips (for fine control) inserted into the 3D cavity, which can tune the frequency down by up to $\sim \SI{3}{\giga\hertz}$. Additionally, we utilize a series of PTFE rods with OHFC tips threaded into the 3D cavity to increase its frequency by up to $\sim \SI{250}{\mega\hertz}$. Choosing a non-superconducting material such as OFHC allows us to generate a magnetic field inside the 3D cavity using an external tuning coil. Additionally, the 3D cavity has two cutouts for optical fiber access for potential microwave-to-optical quantum frequency transduction experiments (see the supplementary material).

To achieve strong electromagnetic coupling between the 3D cavity and the microwave resonator, a microwave resonator with a large dipole moment is desirable. Furthermore, a microwave resonator with high impedance provides~\cite{wu2020microwave, Stockill2022} stronger piezoelectric coupling. These two requirements guide our design choices resulting in a microwave resonator with two large and widely spaced capacitor pads connected by a nanowire kinetic inductor (Fig.~\ref{fig:modes}(c)). To maximize the characteristic impedance of the microwave resonator, we utilize the high-kinetic inductance of a superconducting niobium titanium nitride (NbTiN) film (sputtered by STAR Cryoelectronics, LLC). Because kinetic inductance scales as $L_\text{k}=\mu_\text{o} \lambda_\text{L}^{2}\left(\frac{l}{w t}\right)$, where $\lambda_\text{L}$ is the London penetration depth, and $l, w, t$ represent the length, width, and thickness of a wire, the nanowires are designed to be thin, narrow, and long. The nanowires have a ladder structure to them, allowing us to use the quadratic non-linearity in kinetic inductance, $L_\text{k}\left(I\right) \approx L_\text{k}(0)\left[1+\left(\frac{I}{I^{*}}\right)^{2}\right]$, and wirelessly tune the frequency of the microwave resonator with an external magnetic field~\cite{xu2019frequency}. The microwave resonator is fabricated by patterning \SI{10}{\nano\meter} NbTiN thin film on a high resistivity silicon substrate followed by evaporated aluminum that defines the additional wiring and pads for flip-chip galvanic contacts. At $T\sim \SI{10}{\milli\kelvin}$, one such microwave resonator has a quality factor of $\sim 96\times10^3$ and frequency tuning of the fundamental mode greater than $\SI{100}{\mega\hertz}$ (Fig.~\ref{fig:modes}(d)). 

Thin-film lithium niobate (LN), a strong piezoelectric material, forms the basis of our nanomechanical phononic crystal resonator~\cite{arrangoiz2018coupling}. The nanomechanical resonator consists of a center defect, designed to be at the frequency of interest, placed in a 1D phononic crystal shield. The periodic nature of the 1D phononic crystal opens a complete phononic bandgap in the $ \SI{2.3}-\SI{3.0}{\giga\hertz}$ range that helps confine the mechanical motion to the defect and hence minimizes unwanted mechanical loss~\cite{arrangoiz2018coupling, arrangoiz2019resolving, wollack2021loss}. The full device consists of an array of such nanomechanical resonators (Fig.~\ref{fig:modes}(e)) that are frequency-multiplexed within the bandgap. We fabricate these devices by argon ion milling \SI{250}{\nano\meter} thick thin-film lithium niobate (LN) on a \SI{500}{\micro\meter} thick silicon (Si) handle to define the phononic crystals that are masked by HSQ (FOx-16) resist patterned with electron beam lithography. This is followed by aluminum evaporation and liftoff to create the electrodes and contact pads. The resonators are then released by selectively etching the underlying silicon substrate with XeF$_2$. A room temperature reflection $S_{11}$ measurement (Fig.~\ref{fig:modes}(f)) shows nine dips in the spectrum corresponding to the nine nanomechanical resonators per device with room temperature quality factors $\sim \SI{1800}{}$.

Galvanically connecting the microwave and nanomechanical resonators on the same chip would ideally give the strongest coupling. However, the increased fabrication complexity and reduced yield associated with co-integrating all of the components, makes heterogeneous integration approaches more attractive. Capacitive coupling between a superconducting qubit and a mechanical resonator on two separate chips have previously been used to demonstrate quantum acoustic systems with improved performance~\cite{satzinger2018quantum,wollack2022quantum}. Such a capacitive coupling comes at the cost of diluting the electromechanical coupling through addition of cross-chip coupling and parasitic capacitances to the circuit. This downside is particularly problematic for nanomechanical oscillators which already have extremely small piezoelectric coupling capacitances (see the supplementary material). A heterogeneous approach that still preserves galvanic contacts between two off-chip resonators would greatly simplify fabrication, while still accommodating devices of different scales and footprints. Towards this effort, various flip-chip architectures are developed, where indium bumps serve as the galvanic contacts~\cite{li2021vacuum, lei2020high, foxen2017qubit, rosenberg20173d, kosen2022building}. While the photolithographically patterned indium bumps in these approaches provide controlled bump shapes and small feature sizes, the additional bump fabrication steps, which often include an under-bump metalization layer to act as a diffusion barrier, may not always be compatible with other processes needed for fabricating the devices. In this work, we are interested in a plug-and-play approach that provides all the desired benefits listed above, but also eliminates the need for additional fabrication steps.

\begin{figure}[tb]
\includegraphics[scale=1]{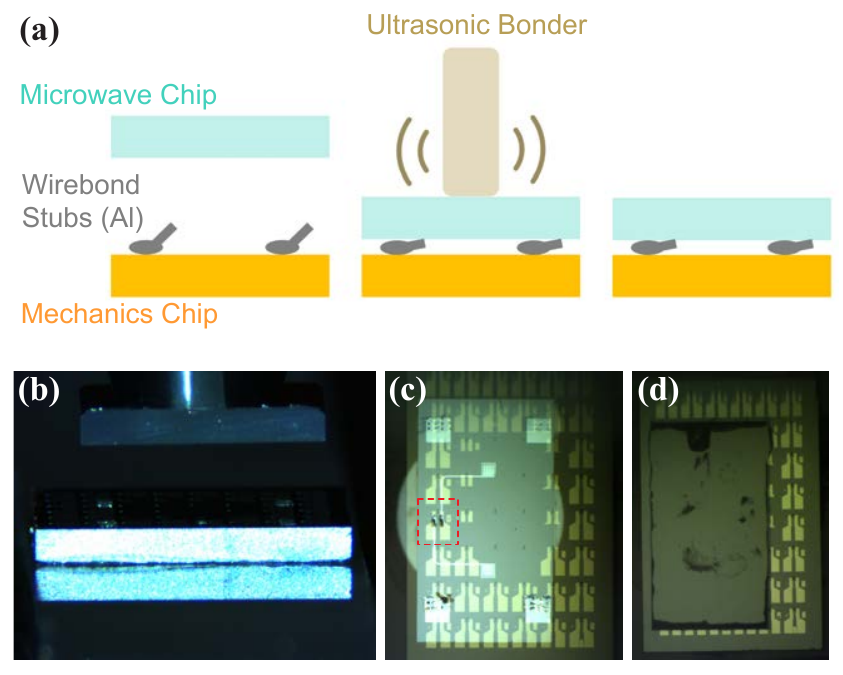}
\caption{\label{fig:fcbb} \textbf{Flip-chip bump bond process.} \textbf{(a)} Schematic of the flip-chip bump bond process flow. Aluminum wirebond stubs are added to the contact pads on the mechanics chip that is placed upright on a vacuum chuck. The inverted and aligned microwave chip is brought in contact with the mechanics chip. Ultrasonic power is applied resulting in the two chips bonding together forming galvanic contacts at the locations of the aluminum stubs. \textbf{(b)} Side view of the two chips before they are brought in contact with each other. \textbf{(c)} Top view showing the alignment process. Alignment of the two chips is performed by overlaying microscopic images of the two chips using a beamsplitter. The highlighted red region corresponds to galvanic contacts between the microwave chip and the mechanics chip. \textbf{(d)} Top view showing the bonded chips. The smaller chip on the top is the microwave chip.}
\end{figure}

The flip-chip approach we developed in this work is shown schematically in Fig.~\ref{fig:fcbb}(a). This approach involves using a wirebonder (West Bond 7476E), configured to place one bond per wire, to place aluminum stubs, which act as the bumps, at the desired galvanic contact locations on the bottom chip (mechanics chip in this case). Additional stubs are added at four corner locations to provide support and stability to the chips post-bonding. We then use a die-bonder tool (FINEPLACER® lambda) where the inverted top chip (microwave chip), held by a vacuum die pick (Fig.~\ref{fig:fcbb}(b)), and the bottom chip (mechanics chip) are aligned to each other using beamsplitter-based optical imaging (Fig.~\ref{fig:fcbb}(c)). We bring the aligned chips in touch with each other and use an ultrasonic technique~\cite{harman2010wire, onuki1987investigation, hizukuri2001dynamic, tani2010development, shuto2015situ} to bond the two chips together (Fig.~\ref{fig:fcbb}(d)). The resulting bonded microwave-mechanics chip pair has an inter-chip separation on the order of $ \SI{10}-\SI{20}{\micro\meter}$. 

\begin{figure}[tb]
\includegraphics[scale=1]{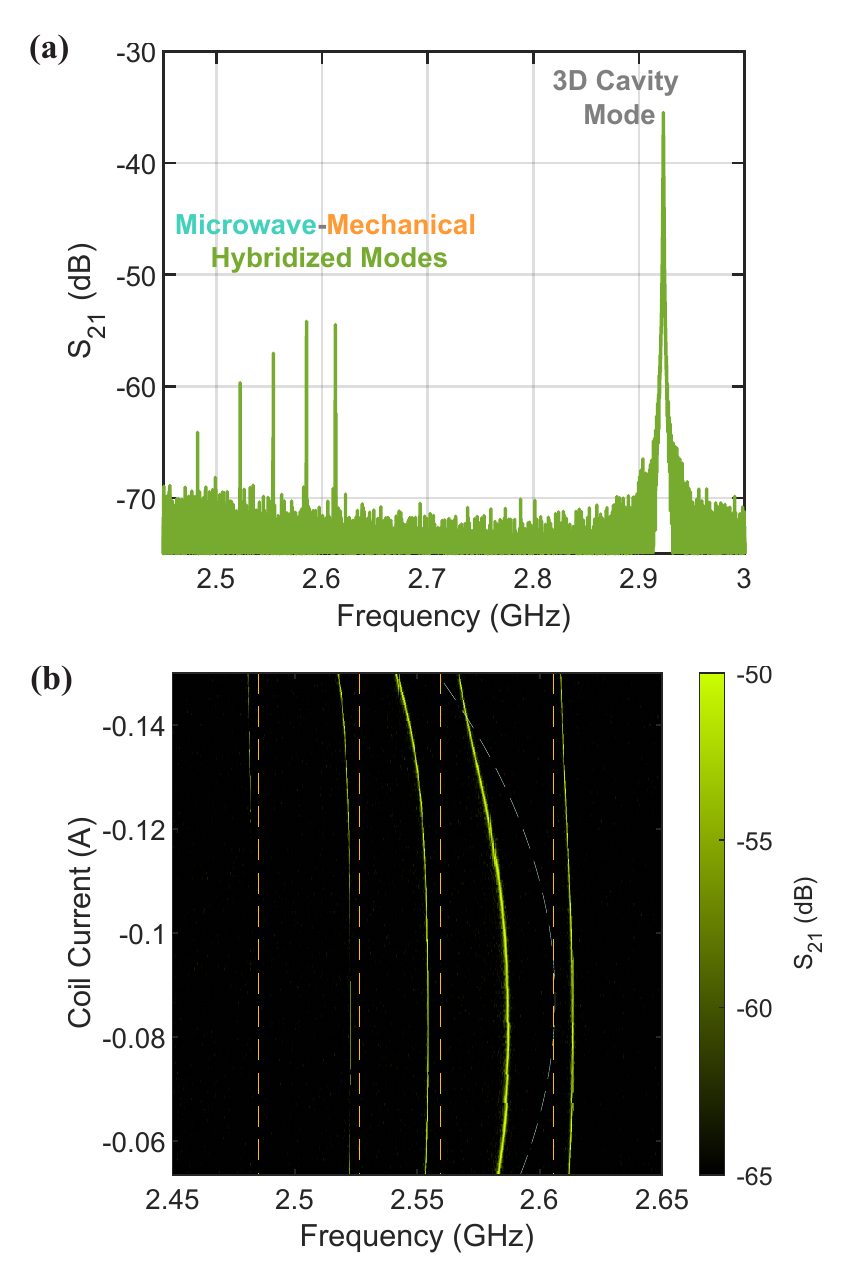}
\caption{\label{fig:coupling} \textbf{Hybridized microwave-mechanical modes.} \textbf{(a)} Cryogenic measurement ($T\sim \SI{10}{\milli\kelvin}$) showing wide transmission spectrum through the 3D cavity. The spectrum shows the tuned down 3D cavity mode at $\omega_\text{c}/2\pi \sim \SI{2.923}{\giga\hertz}$. Because of the close proximity of the 3D cavity mode to the microwave mode, the hybridized microwave-mechanical modes ($ \SI{2.45}-\SI{2.65}{\giga\hertz}$) can be seen as peaks in the transmission spectrum of the 3D cavity. \textbf{(b)} Narrow transmission spectrum as a function of the applied external magnetic field. As the external magnetic field is varied, the microwave mode (blue dashed line) quadratically tunes and crosses into various mechanical modes (yellow dashed lines) resulting in the whole hybridized microwave-mechanical spectrum to shift. From the measurement, a strong microwave-mechanics coupling rate of $\sim \SI{15}{\mega\hertz}$ is extracted (Fig.~\ref{SI-fig:fcbbfit}, Table.~\ref{tab:fcbbparameters}).}
\end{figure}

The microwave-mechanics chip pair is glued onto a sapphire strip that is then inserted and secured into the 3D cavity. The cavity mode is tuned down appropriately to be near the microwave and mechanics mode frequencies before cooling it down in a dilution refrigerator. A transmission measurement $S_{21}$ spectrum at cryogenic temperature ($T\sim \SI{10}{\milli\kelvin}$) shows a prominent peak at $\omega_\text{c}/2\pi = \SI{2.923}{\giga\hertz}$ which corresponds to the cavity mode (Fig.~\ref{fig:coupling}(a)). In close proximity to the cavity peak, we observe a series of peaks ($ \SI{2.45}-\SI{2.65}{\giga\hertz}$), which agrees with the presence of an array of nanomechanical resonators on a single mechanical device in that frequency range. The fact that we observe a series of nanomechanical modes through the cavity transmission spectrum confirms that we have successfully coupled to both the piezoelectric and the electromagnetic modes. 

We apply a magnetic field through an external current-carrying coil (see the supplementary material), and measure the microwave transmission over a smaller frequency window ($\SI{2.45}-\SI{2.65}{\giga\hertz}$) as a function of the applied magnetic field (Fig.~\ref{fig:coupling}(b)). We observe changes in the spectrum as a function of the imposed magnetic field that is consistent with the tuning of the microwave mode. Furthermore, we observe a series of anti-crossings. Taking the trace of this data at three different cuts along the y-axis and fitting it to coupled-mode input-output theory (see the supplementary material) reveals that the five curves are all hybridized microwave-mechanical modes. From the fits of these hybridized modes, we extract the frequencies of four individual mechanical modes (shown as yellow dashed lines) and one microwave mode (shown as blue dashed line) that quadratically tunes with the applied magnetic field. The quality factor of the microwave mode is $\sim 8\times10^3$ and the quality factor of mechanical modes is $\sim 30\times10^3$.  The extracted coupling rate (see the supplementary material) between the microwave mode and the cavity mode is $g_\text{ac}/2\pi = \SI{83.5}{\mega\hertz}$, whereas the extracted coupling rate between the microwave mode and four mechanical modes is $g_\text{ab}/2\pi = \SI{13.6}-\SI{15.3}{\mega\hertz}$, which puts us in the strong-coupling regime.

We compared these results with an alternative integration approach where we arrange a pair of microwave and mechanics chips next to each other in a plane and implement short cross-chip wirebonds between the two. The cryogenic measurement from the cross-chip wirebonded chip pair exhibit similar performance with microwave-mechanics coupling rates between $\SI{12}-\SI{14}{\mega\hertz}$ (see the supplementary material). While this approach apparently provides all the benefits of the flip-chip integration approach presented in this work, such as simplified fabrication, accommodating chips of different scales, and galvanic contacts, the cross-chip wirebond approach suffers from significant uncertainty in the post-integration frequency shift of the microwave mode due to the added stray capacitances that are challenging to control. The sensitivity of this frequency shift to stray capacitances is further exaggerated by the high-impedance nature of the microwave resonator. A lumped element circuit model of the cross-chip wirebond (see the supplementary material) shows that the post-integration frequency shift of the microwave mode is very sensitive to the wirebond length. This poses a serious problem because it becomes difficult to predict where the microwave mode frequency will land after the cross-chip wirebond due to the lack of control over the length and shape of the wirebonds. On the other hand, the flip-chip integration approach presented in this work has a consistent post-integration microwave mode frequency shift of $ 17\pm1\%$. We found that even though the cross-chip wirebond integration method gives comparable coupling rates and quality factors, and has the same flexibility as the flip-chip method discussed in this paper, it is not practical for applications requiring good frequency matching of the modes. It is important to point out that the quality factor of the microwave mode suffers in both these integration approaches i.e. it drops from $\sim\SI{96}{k}$ to $\sim\SI{8}{k}$. This is probably due to the added interface resistance at the wirebond locations.

In conclusion, we have demonstrated strong coupling between a high kinetic inductance superconducting microwave resonator and gigahertz-frequency nanomechanical resonators using an integration approach that utilizes a combination of a 3D cavity and a convenient heterogeneous galvanically bonded flip-chip method. The demonstrated multiple-chip architecture provides flexibility and simplified fabrication, and could potentially enable coupling between a vast variety of quantum systems such as spins with different host materials. Our work demonstrates a way to significantly simplify experiments with hybrid quantum systems that require strong coupling and fairly fine control over system parameters.

The authors would like to thank Rachel G. Gruenke, Nathan R.A. Lee, Kevin K.S. Multani, and Agnetta Y. Cleland for fabrication assistance. This work was primarily supported by the U.S. Army Research Office (ARO) Cross-Quantum Systems Science \& Technology (CQTS) program (Grant No. W911NF-18-1-0103), the National Science Foundation CAREER award No.~ECCS-1941826, the Airforce Office of Scientific Research (AFOSR) (MURI No. FA9550-17-1-0002 led by CUNY), and the David and Lucille Packard Fellowship. Device fabrication was performed at the Stanford Nano Shared Facilities (SNSF) and the Stanford Nanofabrication Facility (SNF), supported by the NSF award ECCS-2026822. T.M. acknowledges support from the National Science Foundation Graduate Research Fellowship Program (grant no. DGE-1656518). A.H.S.-N. acknowledges support via a Sloan Fellowship.  The authors also wish to thank NTT Research and Amazon Web Services Inc. for their financial support. Some of this work was funded by the U.S. Department of Energy through Grant No. DE-AC02-76SF00515 and via the Q-NEXT Center. 



\section*{Data availability}
The data that support the findings of this study are available from the corresponding author upon reasonable request.


\bibliography{ref}

\begin{thebibliography}{46}%
\makeatletter
\providecommand \@ifxundefined [1]{%
 \@ifx{#1\undefined}
}%
\providecommand \@ifnum [1]{%
 \ifnum #1\expandafter \@firstoftwo
 \else \expandafter \@secondoftwo
 \fi
}%
\providecommand \@ifx [1]{%
 \ifx #1\expandafter \@firstoftwo
 \else \expandafter \@secondoftwo
 \fi
}%
\providecommand \natexlab [1]{#1}%
\providecommand \enquote  [1]{``#1''}%
\providecommand \bibnamefont  [1]{#1}%
\providecommand \bibfnamefont [1]{#1}%
\providecommand \citenamefont [1]{#1}%
\providecommand \href@noop [0]{\@secondoftwo}%
\providecommand \href [0]{\begingroup \@sanitize@url \@href}%
\providecommand \@href[1]{\@@startlink{#1}\@@href}%
\providecommand \@@href[1]{\endgroup#1\@@endlink}%
\providecommand \@sanitize@url [0]{\catcode `\\12\catcode `\$12\catcode
  `\&12\catcode `\#12\catcode `\^12\catcode `\_12\catcode `\%12\relax}%
\providecommand \@@startlink[1]{}%
\providecommand \@@endlink[0]{}%
\providecommand \url  [0]{\begingroup\@sanitize@url \@url }%
\providecommand \@url [1]{\endgroup\@href {#1}{\urlprefix }}%
\providecommand \urlprefix  [0]{URL }%
\providecommand \Eprint [0]{\href }%
\providecommand \doibase [0]{http://dx.doi.org/}%
\providecommand \selectlanguage [0]{\@gobble}%
\providecommand \bibinfo  [0]{\@secondoftwo}%
\providecommand \bibfield  [0]{\@secondoftwo}%
\providecommand \translation [1]{[#1]}%
\providecommand \BibitemOpen [0]{}%
\providecommand \bibitemStop [0]{}%
\providecommand \bibitemNoStop [0]{.\EOS\space}%
\providecommand \EOS [0]{\spacefactor3000\relax}%
\providecommand \BibitemShut  [1]{\csname bibitem#1\endcsname}%
\let\auto@bib@innerbib\@empty
\bibitem [{\citenamefont {Clerk}\ \emph {et~al.}(2020)\citenamefont {Clerk},
  \citenamefont {Lehnert}, \citenamefont {Bertet}, \citenamefont {Petta},\ and\
  \citenamefont {Nakamura}}]{Clerk2020}%
  \BibitemOpen
  \bibfield  {author} {\bibinfo {author} {\bibfnamefont {A.~A.}\ \bibnamefont
  {Clerk}}, \bibinfo {author} {\bibfnamefont {K.~W.}\ \bibnamefont {Lehnert}},
  \bibinfo {author} {\bibfnamefont {P.}~\bibnamefont {Bertet}}, \bibinfo
  {author} {\bibfnamefont {J.~R.}\ \bibnamefont {Petta}}, \ and\ \bibinfo
  {author} {\bibfnamefont {Y.}~\bibnamefont {Nakamura}},\ }\href {\doibase
  10.1038/s41567-020-0797-9} {\bibfield  {journal} {\bibinfo  {journal} {Nature
  Physics}\ }\textbf {\bibinfo {volume} {16}},\ \bibinfo {pages} {257}
  (\bibinfo {year} {2020})}\BibitemShut {NoStop}%
\bibitem [{\citenamefont {Mason}\ \emph {et~al.}(2019)\citenamefont {Mason},
  \citenamefont {Chen}, \citenamefont {Rossi}, \citenamefont {Tsaturyan},\ and\
  \citenamefont {Schliesser}}]{Mason2019}%
  \BibitemOpen
  \bibfield  {author} {\bibinfo {author} {\bibfnamefont {D.}~\bibnamefont
  {Mason}}, \bibinfo {author} {\bibfnamefont {J.}~\bibnamefont {Chen}},
  \bibinfo {author} {\bibfnamefont {M.}~\bibnamefont {Rossi}}, \bibinfo
  {author} {\bibfnamefont {Y.}~\bibnamefont {Tsaturyan}}, \ and\ \bibinfo
  {author} {\bibfnamefont {A.}~\bibnamefont {Schliesser}},\ }\href {\doibase
  10.1038/s41567-019-0533-5} {\bibfield  {journal} {\bibinfo  {journal} {Nature
  Physics}\ }\textbf {\bibinfo {volume} {15}},\ \bibinfo {pages} {745}
  (\bibinfo {year} {2019})}\BibitemShut {NoStop}%
\bibitem [{\citenamefont {Pechal}\ \emph {et~al.}(2018)\citenamefont {Pechal},
  \citenamefont {Arrangoiz-Arriola},\ and\ \citenamefont
  {Safavi-Naeini}}]{Pechal2018}%
  \BibitemOpen
  \bibfield  {author} {\bibinfo {author} {\bibfnamefont {M.}~\bibnamefont
  {Pechal}}, \bibinfo {author} {\bibfnamefont {P.}~\bibnamefont
  {Arrangoiz-Arriola}}, \ and\ \bibinfo {author} {\bibfnamefont {A.~H.}\
  \bibnamefont {Safavi-Naeini}},\ }\href {\doibase 10.1088/2058-9565/AADC6C}
  {\bibfield  {journal} {\bibinfo  {journal} {Quantum Science and Technology}\
  }\textbf {\bibinfo {volume} {4}},\ \bibinfo {pages} {015006} (\bibinfo {year}
  {2018})}\BibitemShut {NoStop}%
\bibitem [{\citenamefont {Hann}\ \emph {et~al.}(2019)\citenamefont {Hann},
  \citenamefont {Zou}, \citenamefont {Zhang}, \citenamefont {Chu},
  \citenamefont {Schoelkopf}, \citenamefont {Girvin},\ and\ \citenamefont
  {Jiang}}]{Hann2019}%
  \BibitemOpen
  \bibfield  {author} {\bibinfo {author} {\bibfnamefont {C.~T.}\ \bibnamefont
  {Hann}}, \bibinfo {author} {\bibfnamefont {C.~L.}\ \bibnamefont {Zou}},
  \bibinfo {author} {\bibfnamefont {Y.}~\bibnamefont {Zhang}}, \bibinfo
  {author} {\bibfnamefont {Y.}~\bibnamefont {Chu}}, \bibinfo {author}
  {\bibfnamefont {R.~J.}\ \bibnamefont {Schoelkopf}}, \bibinfo {author}
  {\bibfnamefont {S.~M.}\ \bibnamefont {Girvin}}, \ and\ \bibinfo {author}
  {\bibfnamefont {L.}~\bibnamefont {Jiang}},\ }\href {\doibase
  10.1103/PHYSREVLETT.123.250501/FIGURES/4/MEDIUM} {\bibfield  {journal}
  {\bibinfo  {journal} {Physical Review Letters}\ }\textbf {\bibinfo {volume}
  {123}},\ \bibinfo {pages} {250501} (\bibinfo {year} {2019})}\BibitemShut
  {NoStop}%
\bibitem [{\citenamefont {Jiang}\ \emph {et~al.}(2022)\citenamefont {Jiang},
  \citenamefont {Mayor}, \citenamefont {Malik}, \citenamefont {Van~Laer},
  \citenamefont {McKenna}, \citenamefont {Patel}, \citenamefont {Witmer},\ and\
  \citenamefont {Safavi-Naeini}}]{jiang2022optically}%
  \BibitemOpen
  \bibfield  {author} {\bibinfo {author} {\bibfnamefont {W.}~\bibnamefont
  {Jiang}}, \bibinfo {author} {\bibfnamefont {F.~M.}\ \bibnamefont {Mayor}},
  \bibinfo {author} {\bibfnamefont {S.}~\bibnamefont {Malik}}, \bibinfo
  {author} {\bibfnamefont {R.}~\bibnamefont {Van~Laer}}, \bibinfo {author}
  {\bibfnamefont {T.~P.}\ \bibnamefont {McKenna}}, \bibinfo {author}
  {\bibfnamefont {R.~N.}\ \bibnamefont {Patel}}, \bibinfo {author}
  {\bibfnamefont {J.~D.}\ \bibnamefont {Witmer}}, \ and\ \bibinfo {author}
  {\bibfnamefont {A.~H.}\ \bibnamefont {Safavi-Naeini}},\ }\href@noop {}
  {\bibfield  {journal} {\bibinfo  {journal} {arXiv preprint arXiv:2210.10739}\
  } (\bibinfo {year} {2022})}\BibitemShut {NoStop}%
\bibitem [{\citenamefont {Meesala}\ \emph {et~al.}(2023)\citenamefont
  {Meesala}, \citenamefont {Wood}, \citenamefont {Lake}, \citenamefont
  {Chiappina}, \citenamefont {Zhong}, \citenamefont {Beyer}, \citenamefont
  {Shaw}, \citenamefont {Jiang},\ and\ \citenamefont {Painter}}]{Meesala2023}%
  \BibitemOpen
  \bibfield  {author} {\bibinfo {author} {\bibfnamefont {S.}~\bibnamefont
  {Meesala}}, \bibinfo {author} {\bibfnamefont {S.}~\bibnamefont {Wood}},
  \bibinfo {author} {\bibfnamefont {D.}~\bibnamefont {Lake}}, \bibinfo {author}
  {\bibfnamefont {P.}~\bibnamefont {Chiappina}}, \bibinfo {author}
  {\bibfnamefont {C.}~\bibnamefont {Zhong}}, \bibinfo {author} {\bibfnamefont
  {A.~D.}\ \bibnamefont {Beyer}}, \bibinfo {author} {\bibfnamefont {M.~D.}\
  \bibnamefont {Shaw}}, \bibinfo {author} {\bibfnamefont {L.}~\bibnamefont
  {Jiang}}, \ and\ \bibinfo {author} {\bibfnamefont {O.}~\bibnamefont
  {Painter}},\ }\href@noop {} {\bibfield  {journal} {\bibinfo  {journal} {arXiv
  preprint arXiv:2303.17684}\ } (\bibinfo {year} {2023})}\BibitemShut {NoStop}%
\bibitem [{\citenamefont {Weaver}\ \emph {et~al.}(2022)\citenamefont {Weaver},
  \citenamefont {Duivestein}, \citenamefont {Bernasconi}, \citenamefont
  {Scharmer}, \citenamefont {Lemang}, \citenamefont {van Thiel}, \citenamefont
  {Hijazi}, \citenamefont {Hensen}, \citenamefont {Gr{\"o}blacher},\ and\
  \citenamefont {Stockill}}]{Weaver2022}%
  \BibitemOpen
  \bibfield  {author} {\bibinfo {author} {\bibfnamefont {M.~J.}\ \bibnamefont
  {Weaver}}, \bibinfo {author} {\bibfnamefont {P.}~\bibnamefont {Duivestein}},
  \bibinfo {author} {\bibfnamefont {A.~C.}\ \bibnamefont {Bernasconi}},
  \bibinfo {author} {\bibfnamefont {S.}~\bibnamefont {Scharmer}}, \bibinfo
  {author} {\bibfnamefont {M.}~\bibnamefont {Lemang}}, \bibinfo {author}
  {\bibfnamefont {T.~C.}\ \bibnamefont {van Thiel}}, \bibinfo {author}
  {\bibfnamefont {F.}~\bibnamefont {Hijazi}}, \bibinfo {author} {\bibfnamefont
  {B.}~\bibnamefont {Hensen}}, \bibinfo {author} {\bibfnamefont
  {S.}~\bibnamefont {Gr{\"o}blacher}}, \ and\ \bibinfo {author} {\bibfnamefont
  {R.}~\bibnamefont {Stockill}},\ }\href@noop {} {\bibfield  {journal}
  {\bibinfo  {journal} {arXiv preprint arXiv:2210.15702}\ } (\bibinfo {year}
  {2022})}\BibitemShut {NoStop}%
\bibitem [{\citenamefont {Mirhosseini}\ \emph {et~al.}(2020)\citenamefont
  {Mirhosseini}, \citenamefont {Sipahigil}, \citenamefont {Kalaee},\ and\
  \citenamefont {Painter}}]{mirhosseini2020superconducting}%
  \BibitemOpen
  \bibfield  {author} {\bibinfo {author} {\bibfnamefont {M.}~\bibnamefont
  {Mirhosseini}}, \bibinfo {author} {\bibfnamefont {A.}~\bibnamefont
  {Sipahigil}}, \bibinfo {author} {\bibfnamefont {M.}~\bibnamefont {Kalaee}}, \
  and\ \bibinfo {author} {\bibfnamefont {O.}~\bibnamefont {Painter}},\
  }\href@noop {} {\bibfield  {journal} {\bibinfo  {journal} {Nature}\ }\textbf
  {\bibinfo {volume} {588}},\ \bibinfo {pages} {599} (\bibinfo {year}
  {2020})}\BibitemShut {NoStop}%
\bibitem [{\citenamefont {Jiang}\ \emph {et~al.}(2020)\citenamefont {Jiang},
  \citenamefont {Sarabalis}, \citenamefont {Dahmani}, \citenamefont {Patel},
  \citenamefont {Mayor}, \citenamefont {McKenna}, \citenamefont {Van~Laer},\
  and\ \citenamefont {Safavi-Naeini}}]{jiang2020efficient}%
  \BibitemOpen
  \bibfield  {author} {\bibinfo {author} {\bibfnamefont {W.}~\bibnamefont
  {Jiang}}, \bibinfo {author} {\bibfnamefont {C.~J.}\ \bibnamefont
  {Sarabalis}}, \bibinfo {author} {\bibfnamefont {Y.~D.}\ \bibnamefont
  {Dahmani}}, \bibinfo {author} {\bibfnamefont {R.~N.}\ \bibnamefont {Patel}},
  \bibinfo {author} {\bibfnamefont {F.~M.}\ \bibnamefont {Mayor}}, \bibinfo
  {author} {\bibfnamefont {T.~P.}\ \bibnamefont {McKenna}}, \bibinfo {author}
  {\bibfnamefont {R.}~\bibnamefont {Van~Laer}}, \ and\ \bibinfo {author}
  {\bibfnamefont {A.~H.}\ \bibnamefont {Safavi-Naeini}},\ }\href@noop {}
  {\bibfield  {journal} {\bibinfo  {journal} {Nature communications}\ }\textbf
  {\bibinfo {volume} {11}},\ \bibinfo {pages} {1} (\bibinfo {year}
  {2020})}\BibitemShut {NoStop}%
\bibitem [{\citenamefont {Han}\ \emph {et~al.}(2020)\citenamefont {Han},
  \citenamefont {Fu}, \citenamefont {Zhong}, \citenamefont {Zou}, \citenamefont
  {Xu}, \citenamefont {Sayem}, \citenamefont {Xu}, \citenamefont {Wang},
  \citenamefont {Cheng}, \citenamefont {Jiang} \emph {et~al.}}]{han2020cavity}%
  \BibitemOpen
  \bibfield  {author} {\bibinfo {author} {\bibfnamefont {X.}~\bibnamefont
  {Han}}, \bibinfo {author} {\bibfnamefont {W.}~\bibnamefont {Fu}}, \bibinfo
  {author} {\bibfnamefont {C.}~\bibnamefont {Zhong}}, \bibinfo {author}
  {\bibfnamefont {C.-L.}\ \bibnamefont {Zou}}, \bibinfo {author} {\bibfnamefont
  {Y.}~\bibnamefont {Xu}}, \bibinfo {author} {\bibfnamefont {A.~A.}\
  \bibnamefont {Sayem}}, \bibinfo {author} {\bibfnamefont {M.}~\bibnamefont
  {Xu}}, \bibinfo {author} {\bibfnamefont {S.}~\bibnamefont {Wang}}, \bibinfo
  {author} {\bibfnamefont {R.}~\bibnamefont {Cheng}}, \bibinfo {author}
  {\bibfnamefont {L.}~\bibnamefont {Jiang}},  \emph {et~al.},\ }\href@noop {}
  {\bibfield  {journal} {\bibinfo  {journal} {Nature communications}\ }\textbf
  {\bibinfo {volume} {11}},\ \bibinfo {pages} {1} (\bibinfo {year}
  {2020})}\BibitemShut {NoStop}%
\bibitem [{\citenamefont {Han}\ \emph {et~al.}(2021)\citenamefont {Han},
  \citenamefont {Fu}, \citenamefont {Zou}, \citenamefont {Jiang},\ and\
  \citenamefont {Tang}}]{han2021microwave}%
  \BibitemOpen
  \bibfield  {author} {\bibinfo {author} {\bibfnamefont {X.}~\bibnamefont
  {Han}}, \bibinfo {author} {\bibfnamefont {W.}~\bibnamefont {Fu}}, \bibinfo
  {author} {\bibfnamefont {C.-L.}\ \bibnamefont {Zou}}, \bibinfo {author}
  {\bibfnamefont {L.}~\bibnamefont {Jiang}}, \ and\ \bibinfo {author}
  {\bibfnamefont {H.~X.}\ \bibnamefont {Tang}},\ }\href@noop {} {\bibfield
  {journal} {\bibinfo  {journal} {Optica}\ }\textbf {\bibinfo {volume} {8}},\
  \bibinfo {pages} {1050} (\bibinfo {year} {2021})}\BibitemShut {NoStop}%
\bibitem [{\citenamefont {Blais}\ \emph {et~al.}(2004)\citenamefont {Blais},
  \citenamefont {Huang}, \citenamefont {Wallraff}, \citenamefont {Girvin},\
  and\ \citenamefont {Schoelkopf}}]{Blais2004}%
  \BibitemOpen
  \bibfield  {author} {\bibinfo {author} {\bibfnamefont {A.}~\bibnamefont
  {Blais}}, \bibinfo {author} {\bibfnamefont {R.~S.}\ \bibnamefont {Huang}},
  \bibinfo {author} {\bibfnamefont {A.}~\bibnamefont {Wallraff}}, \bibinfo
  {author} {\bibfnamefont {S.~M.}\ \bibnamefont {Girvin}}, \ and\ \bibinfo
  {author} {\bibfnamefont {R.~J.}\ \bibnamefont {Schoelkopf}},\ }\href
  {\doibase 10.1103/PHYSREVA.69.062320/FIGURES/9/MEDIUM} {\bibfield  {journal}
  {\bibinfo  {journal} {Physical Review A - Atomic, Molecular, and Optical
  Physics}\ }\textbf {\bibinfo {volume} {69}},\ \bibinfo {pages} {062320}
  (\bibinfo {year} {2004})}\BibitemShut {NoStop}%
\bibitem [{\citenamefont {Wallraff}\ \emph {et~al.}(2004)\citenamefont
  {Wallraff}, \citenamefont {Schuster}, \citenamefont {Blais}, \citenamefont
  {Frunzio}, \citenamefont {Huang}, \citenamefont {Majer}, \citenamefont
  {Kumar}, \citenamefont {Girvin},\ and\ \citenamefont
  {Schoelkopf}}]{Wallraff2004}%
  \BibitemOpen
  \bibfield  {author} {\bibinfo {author} {\bibfnamefont {A.}~\bibnamefont
  {Wallraff}}, \bibinfo {author} {\bibfnamefont {D.~I.}\ \bibnamefont
  {Schuster}}, \bibinfo {author} {\bibfnamefont {A.}~\bibnamefont {Blais}},
  \bibinfo {author} {\bibfnamefont {L.}~\bibnamefont {Frunzio}}, \bibinfo
  {author} {\bibfnamefont {R.~S.}\ \bibnamefont {Huang}}, \bibinfo {author}
  {\bibfnamefont {J.}~\bibnamefont {Majer}}, \bibinfo {author} {\bibfnamefont
  {S.}~\bibnamefont {Kumar}}, \bibinfo {author} {\bibfnamefont {S.~M.}\
  \bibnamefont {Girvin}}, \ and\ \bibinfo {author} {\bibfnamefont {R.~J.}\
  \bibnamefont {Schoelkopf}},\ }\href {\doibase 10.1038/nature02851} {\bibfield
   {journal} {\bibinfo  {journal} {Nature}\ }\textbf {\bibinfo {volume}
  {431}},\ \bibinfo {pages} {162} (\bibinfo {year} {2004})}\BibitemShut
  {NoStop}%
\bibitem [{\citenamefont {Paik}\ \emph {et~al.}(2011)\citenamefont {Paik},
  \citenamefont {Schuster}, \citenamefont {Bishop}, \citenamefont {Kirchmair},
  \citenamefont {Catelani}, \citenamefont {Sears}, \citenamefont {Johnson},
  \citenamefont {Reagor}, \citenamefont {Frunzio}, \citenamefont {Glazman},
  \citenamefont {Girvin}, \citenamefont {Devoret},\ and\ \citenamefont
  {Schoelkopf}}]{Paik2011}%
  \BibitemOpen
  \bibfield  {author} {\bibinfo {author} {\bibfnamefont {H.}~\bibnamefont
  {Paik}}, \bibinfo {author} {\bibfnamefont {D.~I.}\ \bibnamefont {Schuster}},
  \bibinfo {author} {\bibfnamefont {L.~S.}\ \bibnamefont {Bishop}}, \bibinfo
  {author} {\bibfnamefont {G.}~\bibnamefont {Kirchmair}}, \bibinfo {author}
  {\bibfnamefont {G.}~\bibnamefont {Catelani}}, \bibinfo {author}
  {\bibfnamefont {A.~P.}\ \bibnamefont {Sears}}, \bibinfo {author}
  {\bibfnamefont {B.~R.}\ \bibnamefont {Johnson}}, \bibinfo {author}
  {\bibfnamefont {M.~J.}\ \bibnamefont {Reagor}}, \bibinfo {author}
  {\bibfnamefont {L.}~\bibnamefont {Frunzio}}, \bibinfo {author} {\bibfnamefont
  {L.~I.}\ \bibnamefont {Glazman}}, \bibinfo {author} {\bibfnamefont {S.~M.}\
  \bibnamefont {Girvin}}, \bibinfo {author} {\bibfnamefont {M.~H.}\
  \bibnamefont {Devoret}}, \ and\ \bibinfo {author} {\bibfnamefont {R.~J.}\
  \bibnamefont {Schoelkopf}},\ }\href {\doibase 10.1103/PhysRevLett.107.240501}
  {\bibfield  {journal} {\bibinfo  {journal} {Physical Review Letters}\
  }\textbf {\bibinfo {volume} {107}},\ \bibinfo {pages} {240501} (\bibinfo
  {year} {2011})}\BibitemShut {NoStop}%
\bibitem [{\citenamefont {Wang}\ \emph {et~al.}(2014)\citenamefont {Wang},
  \citenamefont {Gao}, \citenamefont {Pop}, \citenamefont {Vool}, \citenamefont
  {Axline}, \citenamefont {Brecht}, \citenamefont {Heeres}, \citenamefont
  {Frunzio}, \citenamefont {Devoret}, \citenamefont {Catelani}, \citenamefont
  {Glazman},\ and\ \citenamefont {Schoelkopf}}]{Wang2014}%
  \BibitemOpen
  \bibfield  {author} {\bibinfo {author} {\bibfnamefont {C.}~\bibnamefont
  {Wang}}, \bibinfo {author} {\bibfnamefont {Y.~Y.}\ \bibnamefont {Gao}},
  \bibinfo {author} {\bibfnamefont {I.~M.}\ \bibnamefont {Pop}}, \bibinfo
  {author} {\bibfnamefont {U.}~\bibnamefont {Vool}}, \bibinfo {author}
  {\bibfnamefont {C.}~\bibnamefont {Axline}}, \bibinfo {author} {\bibfnamefont
  {T.}~\bibnamefont {Brecht}}, \bibinfo {author} {\bibfnamefont {R.~W.}\
  \bibnamefont {Heeres}}, \bibinfo {author} {\bibfnamefont {L.}~\bibnamefont
  {Frunzio}}, \bibinfo {author} {\bibfnamefont {M.~H.}\ \bibnamefont
  {Devoret}}, \bibinfo {author} {\bibfnamefont {G.}~\bibnamefont {Catelani}},
  \bibinfo {author} {\bibfnamefont {L.~I.}\ \bibnamefont {Glazman}}, \ and\
  \bibinfo {author} {\bibfnamefont {R.~J.}\ \bibnamefont {Schoelkopf}},\ }\href
  {\doibase 10.1038/ncomms6836} {\bibfield  {journal} {\bibinfo  {journal}
  {Nature Communications}\ }\textbf {\bibinfo {volume} {5}},\ \bibinfo {pages}
  {5836} (\bibinfo {year} {2014})}\BibitemShut {NoStop}%
\bibitem [{\citenamefont {Satzinger}\ \emph {et~al.}(2019)\citenamefont
  {Satzinger}, \citenamefont {Conner}, \citenamefont {Bienfait}, \citenamefont
  {Chang}, \citenamefont {Chou}, \citenamefont {Cleland}, \citenamefont
  {Dumur}, \citenamefont {Grebel}, \citenamefont {Peairs}, \citenamefont
  {Povey} \emph {et~al.}}]{satzinger2019simple}%
  \BibitemOpen
  \bibfield  {author} {\bibinfo {author} {\bibfnamefont {K.}~\bibnamefont
  {Satzinger}}, \bibinfo {author} {\bibfnamefont {C.}~\bibnamefont {Conner}},
  \bibinfo {author} {\bibfnamefont {A.}~\bibnamefont {Bienfait}}, \bibinfo
  {author} {\bibfnamefont {H.-S.}\ \bibnamefont {Chang}}, \bibinfo {author}
  {\bibfnamefont {M.-H.}\ \bibnamefont {Chou}}, \bibinfo {author}
  {\bibfnamefont {A.}~\bibnamefont {Cleland}}, \bibinfo {author} {\bibfnamefont
  {{\'E}.}~\bibnamefont {Dumur}}, \bibinfo {author} {\bibfnamefont
  {J.}~\bibnamefont {Grebel}}, \bibinfo {author} {\bibfnamefont
  {G.}~\bibnamefont {Peairs}}, \bibinfo {author} {\bibfnamefont
  {R.}~\bibnamefont {Povey}},  \emph {et~al.},\ }\href@noop {} {\bibfield
  {journal} {\bibinfo  {journal} {Applied Physics Letters}\ }\textbf {\bibinfo
  {volume} {114}},\ \bibinfo {pages} {173501} (\bibinfo {year}
  {2019})}\BibitemShut {NoStop}%
\bibitem [{\citenamefont {Conner}\ \emph {et~al.}(2021)\citenamefont {Conner},
  \citenamefont {Bienfait}, \citenamefont {Chang}, \citenamefont {Chou},
  \citenamefont {Dumur}, \citenamefont {Grebel}, \citenamefont {Peairs},
  \citenamefont {Povey}, \citenamefont {Yan}, \citenamefont {Zhong} \emph
  {et~al.}}]{conner2021superconducting}%
  \BibitemOpen
  \bibfield  {author} {\bibinfo {author} {\bibfnamefont {C.}~\bibnamefont
  {Conner}}, \bibinfo {author} {\bibfnamefont {A.}~\bibnamefont {Bienfait}},
  \bibinfo {author} {\bibfnamefont {H.-S.}\ \bibnamefont {Chang}}, \bibinfo
  {author} {\bibfnamefont {M.-H.}\ \bibnamefont {Chou}}, \bibinfo {author}
  {\bibfnamefont {{\'E}.}~\bibnamefont {Dumur}}, \bibinfo {author}
  {\bibfnamefont {J.}~\bibnamefont {Grebel}}, \bibinfo {author} {\bibfnamefont
  {G.}~\bibnamefont {Peairs}}, \bibinfo {author} {\bibfnamefont
  {R.}~\bibnamefont {Povey}}, \bibinfo {author} {\bibfnamefont
  {H.}~\bibnamefont {Yan}}, \bibinfo {author} {\bibfnamefont {Y.}~\bibnamefont
  {Zhong}},  \emph {et~al.},\ }\href@noop {} {\bibfield  {journal} {\bibinfo
  {journal} {Applied Physics Letters}\ }\textbf {\bibinfo {volume} {118}},\
  \bibinfo {pages} {232602} (\bibinfo {year} {2021})}\BibitemShut {NoStop}%
\bibitem [{\citenamefont {Chu}\ \emph {et~al.}(2017)\citenamefont {Chu},
  \citenamefont {Kharel}, \citenamefont {Renninger}, \citenamefont {Burkhart},
  \citenamefont {Frunzio}, \citenamefont {Rakich},\ and\ \citenamefont
  {Schoelkopf}}]{chu2017quantum}%
  \BibitemOpen
  \bibfield  {author} {\bibinfo {author} {\bibfnamefont {Y.}~\bibnamefont
  {Chu}}, \bibinfo {author} {\bibfnamefont {P.}~\bibnamefont {Kharel}},
  \bibinfo {author} {\bibfnamefont {W.~H.}\ \bibnamefont {Renninger}}, \bibinfo
  {author} {\bibfnamefont {L.~D.}\ \bibnamefont {Burkhart}}, \bibinfo {author}
  {\bibfnamefont {L.}~\bibnamefont {Frunzio}}, \bibinfo {author} {\bibfnamefont
  {P.~T.}\ \bibnamefont {Rakich}}, \ and\ \bibinfo {author} {\bibfnamefont
  {R.~J.}\ \bibnamefont {Schoelkopf}},\ }\href@noop {} {\bibfield  {journal}
  {\bibinfo  {journal} {Science}\ }\textbf {\bibinfo {volume} {358}},\ \bibinfo
  {pages} {199} (\bibinfo {year} {2017})}\BibitemShut {NoStop}%
\bibitem [{\citenamefont {Chu}\ \emph {et~al.}(2018)\citenamefont {Chu},
  \citenamefont {Kharel}, \citenamefont {Yoon}, \citenamefont {Frunzio},
  \citenamefont {Rakich},\ and\ \citenamefont {Schoelkopf}}]{chu2018creation}%
  \BibitemOpen
  \bibfield  {author} {\bibinfo {author} {\bibfnamefont {Y.}~\bibnamefont
  {Chu}}, \bibinfo {author} {\bibfnamefont {P.}~\bibnamefont {Kharel}},
  \bibinfo {author} {\bibfnamefont {T.}~\bibnamefont {Yoon}}, \bibinfo {author}
  {\bibfnamefont {L.}~\bibnamefont {Frunzio}}, \bibinfo {author} {\bibfnamefont
  {P.~T.}\ \bibnamefont {Rakich}}, \ and\ \bibinfo {author} {\bibfnamefont
  {R.~J.}\ \bibnamefont {Schoelkopf}},\ }\href@noop {} {\bibfield  {journal}
  {\bibinfo  {journal} {Nature}\ }\textbf {\bibinfo {volume} {563}},\ \bibinfo
  {pages} {666} (\bibinfo {year} {2018})}\BibitemShut {NoStop}%
\bibitem [{\citenamefont {Arrangoiz-Arriola}\ \emph {et~al.}(2018)\citenamefont
  {Arrangoiz-Arriola}, \citenamefont {Wollack}, \citenamefont {Pechal},
  \citenamefont {Witmer}, \citenamefont {Hill},\ and\ \citenamefont
  {Safavi-Naeini}}]{arrangoiz2018coupling}%
  \BibitemOpen
  \bibfield  {author} {\bibinfo {author} {\bibfnamefont {P.}~\bibnamefont
  {Arrangoiz-Arriola}}, \bibinfo {author} {\bibfnamefont {E.~A.}\ \bibnamefont
  {Wollack}}, \bibinfo {author} {\bibfnamefont {M.}~\bibnamefont {Pechal}},
  \bibinfo {author} {\bibfnamefont {J.~D.}\ \bibnamefont {Witmer}}, \bibinfo
  {author} {\bibfnamefont {J.~T.}\ \bibnamefont {Hill}}, \ and\ \bibinfo
  {author} {\bibfnamefont {A.~H.}\ \bibnamefont {Safavi-Naeini}},\ }\href@noop
  {} {\bibfield  {journal} {\bibinfo  {journal} {Physical Review X}\ }\textbf
  {\bibinfo {volume} {8}},\ \bibinfo {pages} {031007} (\bibinfo {year}
  {2018})}\BibitemShut {NoStop}%
\bibitem [{\citenamefont {Satzinger}\ \emph {et~al.}(2018)\citenamefont
  {Satzinger}, \citenamefont {Zhong}, \citenamefont {Chang}, \citenamefont
  {Peairs}, \citenamefont {Bienfait}, \citenamefont {Chou}, \citenamefont
  {Cleland}, \citenamefont {Conner}, \citenamefont {Dumur}, \citenamefont
  {Grebel} \emph {et~al.}}]{satzinger2018quantum}%
  \BibitemOpen
  \bibfield  {author} {\bibinfo {author} {\bibfnamefont {K.~J.}\ \bibnamefont
  {Satzinger}}, \bibinfo {author} {\bibfnamefont {Y.}~\bibnamefont {Zhong}},
  \bibinfo {author} {\bibfnamefont {H.-S.}\ \bibnamefont {Chang}}, \bibinfo
  {author} {\bibfnamefont {G.~A.}\ \bibnamefont {Peairs}}, \bibinfo {author}
  {\bibfnamefont {A.}~\bibnamefont {Bienfait}}, \bibinfo {author}
  {\bibfnamefont {M.-H.}\ \bibnamefont {Chou}}, \bibinfo {author}
  {\bibfnamefont {A.}~\bibnamefont {Cleland}}, \bibinfo {author} {\bibfnamefont
  {C.~R.}\ \bibnamefont {Conner}}, \bibinfo {author} {\bibfnamefont
  {{\'E}.}~\bibnamefont {Dumur}}, \bibinfo {author} {\bibfnamefont
  {J.}~\bibnamefont {Grebel}},  \emph {et~al.},\ }\href@noop {} {\bibfield
  {journal} {\bibinfo  {journal} {Nature}\ }\textbf {\bibinfo {volume} {563}},\
  \bibinfo {pages} {661} (\bibinfo {year} {2018})}\BibitemShut {NoStop}%
\bibitem [{\citenamefont {Arrangoiz-Arriola}\ \emph {et~al.}(2019)\citenamefont
  {Arrangoiz-Arriola}, \citenamefont {Wollack}, \citenamefont {Wang},
  \citenamefont {Pechal}, \citenamefont {Jiang}, \citenamefont {McKenna},
  \citenamefont {Witmer}, \citenamefont {Van~Laer},\ and\ \citenamefont
  {Safavi-Naeini}}]{arrangoiz2019resolving}%
  \BibitemOpen
  \bibfield  {author} {\bibinfo {author} {\bibfnamefont {P.}~\bibnamefont
  {Arrangoiz-Arriola}}, \bibinfo {author} {\bibfnamefont {E.~A.}\ \bibnamefont
  {Wollack}}, \bibinfo {author} {\bibfnamefont {Z.}~\bibnamefont {Wang}},
  \bibinfo {author} {\bibfnamefont {M.}~\bibnamefont {Pechal}}, \bibinfo
  {author} {\bibfnamefont {W.}~\bibnamefont {Jiang}}, \bibinfo {author}
  {\bibfnamefont {T.~P.}\ \bibnamefont {McKenna}}, \bibinfo {author}
  {\bibfnamefont {J.~D.}\ \bibnamefont {Witmer}}, \bibinfo {author}
  {\bibfnamefont {R.}~\bibnamefont {Van~Laer}}, \ and\ \bibinfo {author}
  {\bibfnamefont {A.~H.}\ \bibnamefont {Safavi-Naeini}},\ }\href@noop {}
  {\bibfield  {journal} {\bibinfo  {journal} {Nature}\ }\textbf {\bibinfo
  {volume} {571}},\ \bibinfo {pages} {537} (\bibinfo {year}
  {2019})}\BibitemShut {NoStop}%
\bibitem [{\citenamefont {Sletten}\ \emph {et~al.}(2019)\citenamefont
  {Sletten}, \citenamefont {Moores}, \citenamefont {Viennot},\ and\
  \citenamefont {Lehnert}}]{sletten2019resolving}%
  \BibitemOpen
  \bibfield  {author} {\bibinfo {author} {\bibfnamefont {L.~R.}\ \bibnamefont
  {Sletten}}, \bibinfo {author} {\bibfnamefont {B.~A.}\ \bibnamefont {Moores}},
  \bibinfo {author} {\bibfnamefont {J.~J.}\ \bibnamefont {Viennot}}, \ and\
  \bibinfo {author} {\bibfnamefont {K.~W.}\ \bibnamefont {Lehnert}},\
  }\href@noop {} {\bibfield  {journal} {\bibinfo  {journal} {Physical Review
  X}\ }\textbf {\bibinfo {volume} {9}},\ \bibinfo {pages} {021056} (\bibinfo
  {year} {2019})}\BibitemShut {NoStop}%
\bibitem [{\citenamefont {Peterson}\ \emph {et~al.}(2019)\citenamefont
  {Peterson}, \citenamefont {Kotler}, \citenamefont {Lecocq}, \citenamefont
  {Cicak}, \citenamefont {Jin}, \citenamefont {Simmonds}, \citenamefont
  {Aumentado},\ and\ \citenamefont {Teufel}}]{peterson2019ultrastrong}%
  \BibitemOpen
  \bibfield  {author} {\bibinfo {author} {\bibfnamefont {G.}~\bibnamefont
  {Peterson}}, \bibinfo {author} {\bibfnamefont {S.}~\bibnamefont {Kotler}},
  \bibinfo {author} {\bibfnamefont {F.}~\bibnamefont {Lecocq}}, \bibinfo
  {author} {\bibfnamefont {K.}~\bibnamefont {Cicak}}, \bibinfo {author}
  {\bibfnamefont {X.}~\bibnamefont {Jin}}, \bibinfo {author} {\bibfnamefont
  {R.}~\bibnamefont {Simmonds}}, \bibinfo {author} {\bibfnamefont
  {J.}~\bibnamefont {Aumentado}}, \ and\ \bibinfo {author} {\bibfnamefont
  {J.}~\bibnamefont {Teufel}},\ }\href@noop {} {\bibfield  {journal} {\bibinfo
  {journal} {Physical Review Letters}\ }\textbf {\bibinfo {volume} {123}},\
  \bibinfo {pages} {247701} (\bibinfo {year} {2019})}\BibitemShut {NoStop}%
\bibitem [{\citenamefont {Wollack}\ \emph {et~al.}(2022)\citenamefont
  {Wollack}, \citenamefont {Cleland}, \citenamefont {Gruenke}, \citenamefont
  {Wang}, \citenamefont {Arrangoiz-Arriola},\ and\ \citenamefont
  {Safavi-Naeini}}]{wollack2022quantum}%
  \BibitemOpen
  \bibfield  {author} {\bibinfo {author} {\bibfnamefont {E.~A.}\ \bibnamefont
  {Wollack}}, \bibinfo {author} {\bibfnamefont {A.~Y.}\ \bibnamefont
  {Cleland}}, \bibinfo {author} {\bibfnamefont {R.~G.}\ \bibnamefont
  {Gruenke}}, \bibinfo {author} {\bibfnamefont {Z.}~\bibnamefont {Wang}},
  \bibinfo {author} {\bibfnamefont {P.}~\bibnamefont {Arrangoiz-Arriola}}, \
  and\ \bibinfo {author} {\bibfnamefont {A.~H.}\ \bibnamefont
  {Safavi-Naeini}},\ }\href@noop {} {\bibfield  {journal} {\bibinfo  {journal}
  {Nature}\ }\textbf {\bibinfo {volume} {604}},\ \bibinfo {pages} {463}
  (\bibinfo {year} {2022})}\BibitemShut {NoStop}%
\bibitem [{\citenamefont {Bienfait}\ \emph {et~al.}(2019)\citenamefont
  {Bienfait}, \citenamefont {Satzinger}, \citenamefont {Zhong}, \citenamefont
  {Chang}, \citenamefont {Chou}, \citenamefont {Conner}, \citenamefont {Dumur},
  \citenamefont {Grebel}, \citenamefont {Peairs}, \citenamefont {Povey} \emph
  {et~al.}}]{bienfait2019phonon}%
  \BibitemOpen
  \bibfield  {author} {\bibinfo {author} {\bibfnamefont {A.}~\bibnamefont
  {Bienfait}}, \bibinfo {author} {\bibfnamefont {K.~J.}\ \bibnamefont
  {Satzinger}}, \bibinfo {author} {\bibfnamefont {Y.}~\bibnamefont {Zhong}},
  \bibinfo {author} {\bibfnamefont {H.-S.}\ \bibnamefont {Chang}}, \bibinfo
  {author} {\bibfnamefont {M.-H.}\ \bibnamefont {Chou}}, \bibinfo {author}
  {\bibfnamefont {C.~R.}\ \bibnamefont {Conner}}, \bibinfo {author}
  {\bibfnamefont {{\'E}.}~\bibnamefont {Dumur}}, \bibinfo {author}
  {\bibfnamefont {J.}~\bibnamefont {Grebel}}, \bibinfo {author} {\bibfnamefont
  {G.~A.}\ \bibnamefont {Peairs}}, \bibinfo {author} {\bibfnamefont {R.~G.}\
  \bibnamefont {Povey}},  \emph {et~al.},\ }\href@noop {} {\bibfield  {journal}
  {\bibinfo  {journal} {Science}\ }\textbf {\bibinfo {volume} {364}},\ \bibinfo
  {pages} {368} (\bibinfo {year} {2019})}\BibitemShut {NoStop}%
\bibitem [{\citenamefont {Chakram}\ \emph {et~al.}(2021)\citenamefont
  {Chakram}, \citenamefont {Oriani}, \citenamefont {Naik}, \citenamefont
  {Dixit}, \citenamefont {He}, \citenamefont {Agrawal}, \citenamefont {Kwon},\
  and\ \citenamefont {Schuster}}]{chakram2021seamless}%
  \BibitemOpen
  \bibfield  {author} {\bibinfo {author} {\bibfnamefont {S.}~\bibnamefont
  {Chakram}}, \bibinfo {author} {\bibfnamefont {A.~E.}\ \bibnamefont {Oriani}},
  \bibinfo {author} {\bibfnamefont {R.~K.}\ \bibnamefont {Naik}}, \bibinfo
  {author} {\bibfnamefont {A.~V.}\ \bibnamefont {Dixit}}, \bibinfo {author}
  {\bibfnamefont {K.}~\bibnamefont {He}}, \bibinfo {author} {\bibfnamefont
  {A.}~\bibnamefont {Agrawal}}, \bibinfo {author} {\bibfnamefont
  {H.}~\bibnamefont {Kwon}}, \ and\ \bibinfo {author} {\bibfnamefont {D.~I.}\
  \bibnamefont {Schuster}},\ }\href@noop {} {\bibfield  {journal} {\bibinfo
  {journal} {Physical Review Letters}\ }\textbf {\bibinfo {volume} {127}},\
  \bibinfo {pages} {107701} (\bibinfo {year} {2021})}\BibitemShut {NoStop}%
\bibitem [{\citenamefont {Suleymanzade}\ \emph {et~al.}(2020)\citenamefont
  {Suleymanzade}, \citenamefont {Anferov}, \citenamefont {Stone}, \citenamefont
  {Naik}, \citenamefont {Oriani}, \citenamefont {Simon},\ and\ \citenamefont
  {Schuster}}]{Suleymanzade2020}%
  \BibitemOpen
  \bibfield  {author} {\bibinfo {author} {\bibfnamefont {A.}~\bibnamefont
  {Suleymanzade}}, \bibinfo {author} {\bibfnamefont {A.}~\bibnamefont
  {Anferov}}, \bibinfo {author} {\bibfnamefont {M.}~\bibnamefont {Stone}},
  \bibinfo {author} {\bibfnamefont {R.~K.}\ \bibnamefont {Naik}}, \bibinfo
  {author} {\bibfnamefont {A.}~\bibnamefont {Oriani}}, \bibinfo {author}
  {\bibfnamefont {J.}~\bibnamefont {Simon}}, \ and\ \bibinfo {author}
  {\bibfnamefont {D.}~\bibnamefont {Schuster}},\ }\href {\doibase
  10.1063/1.5137900} {\bibfield  {journal} {\bibinfo  {journal} {Applied
  Physics Letters}\ }\textbf {\bibinfo {volume} {116}},\ \bibinfo {pages}
  {104001} (\bibinfo {year} {2020})}\BibitemShut {NoStop}%
\bibitem [{\citenamefont {Haroche}\ \emph {et~al.}(2020)\citenamefont
  {Haroche}, \citenamefont {Brune},\ and\ \citenamefont
  {Raimond}}]{Haroche2020}%
  \BibitemOpen
  \bibfield  {author} {\bibinfo {author} {\bibfnamefont {S.}~\bibnamefont
  {Haroche}}, \bibinfo {author} {\bibfnamefont {M.}~\bibnamefont {Brune}}, \
  and\ \bibinfo {author} {\bibfnamefont {J.~M.}\ \bibnamefont {Raimond}},\
  }\href {\doibase 10.1038/s41567-020-0812-1} {\bibfield  {journal} {\bibinfo
  {journal} {Nature Physics}\ }\textbf {\bibinfo {volume} {16}},\ \bibinfo
  {pages} {243} (\bibinfo {year} {2020})}\BibitemShut {NoStop}%
\bibitem [{\citenamefont {Wu}\ \emph {et~al.}(2020)\citenamefont {Wu},
  \citenamefont {Zeuthen}, \citenamefont {Balram},\ and\ \citenamefont
  {Srinivasan}}]{wu2020microwave}%
  \BibitemOpen
  \bibfield  {author} {\bibinfo {author} {\bibfnamefont {M.}~\bibnamefont
  {Wu}}, \bibinfo {author} {\bibfnamefont {E.}~\bibnamefont {Zeuthen}},
  \bibinfo {author} {\bibfnamefont {K.~C.}\ \bibnamefont {Balram}}, \ and\
  \bibinfo {author} {\bibfnamefont {K.}~\bibnamefont {Srinivasan}},\
  }\href@noop {} {\bibfield  {journal} {\bibinfo  {journal} {Physical Review
  Applied}\ }\textbf {\bibinfo {volume} {13}},\ \bibinfo {pages} {014027}
  (\bibinfo {year} {2020})}\BibitemShut {NoStop}%
\bibitem [{\citenamefont {Stockill}\ \emph {et~al.}(2022)\citenamefont
  {Stockill}, \citenamefont {Forsch}, \citenamefont {Hijazi}, \citenamefont
  {Beaudoin}, \citenamefont {Pantzas}, \citenamefont {Sagnes}, \citenamefont
  {Braive},\ and\ \citenamefont {Gröblacher}}]{Stockill2022}%
  \BibitemOpen
  \bibfield  {author} {\bibinfo {author} {\bibfnamefont {R.}~\bibnamefont
  {Stockill}}, \bibinfo {author} {\bibfnamefont {M.}~\bibnamefont {Forsch}},
  \bibinfo {author} {\bibfnamefont {F.}~\bibnamefont {Hijazi}}, \bibinfo
  {author} {\bibfnamefont {G.}~\bibnamefont {Beaudoin}}, \bibinfo {author}
  {\bibfnamefont {K.}~\bibnamefont {Pantzas}}, \bibinfo {author} {\bibfnamefont
  {I.}~\bibnamefont {Sagnes}}, \bibinfo {author} {\bibfnamefont
  {R.}~\bibnamefont {Braive}}, \ and\ \bibinfo {author} {\bibfnamefont
  {S.}~\bibnamefont {Gröblacher}},\ }\href {\doibase
  10.1038/s41467-022-34338-x} {\bibfield  {journal} {\bibinfo  {journal}
  {Nature Communications}\ }\textbf {\bibinfo {volume} {13}},\ \bibinfo {pages}
  {1} (\bibinfo {year} {2022})}\BibitemShut {NoStop}%
\bibitem [{\citenamefont {Xu}\ \emph {et~al.}(2019)\citenamefont {Xu},
  \citenamefont {Han}, \citenamefont {Fu}, \citenamefont {Zou},\ and\
  \citenamefont {Tang}}]{xu2019frequency}%
  \BibitemOpen
  \bibfield  {author} {\bibinfo {author} {\bibfnamefont {M.}~\bibnamefont
  {Xu}}, \bibinfo {author} {\bibfnamefont {X.}~\bibnamefont {Han}}, \bibinfo
  {author} {\bibfnamefont {W.}~\bibnamefont {Fu}}, \bibinfo {author}
  {\bibfnamefont {C.-L.}\ \bibnamefont {Zou}}, \ and\ \bibinfo {author}
  {\bibfnamefont {H.~X.}\ \bibnamefont {Tang}},\ }\href@noop {} {\bibfield
  {journal} {\bibinfo  {journal} {Applied Physics Letters}\ }\textbf {\bibinfo
  {volume} {114}},\ \bibinfo {pages} {192601} (\bibinfo {year}
  {2019})}\BibitemShut {NoStop}%
\bibitem [{\citenamefont {Wollack}\ \emph {et~al.}(2021)\citenamefont
  {Wollack}, \citenamefont {Cleland}, \citenamefont {Arrangoiz-Arriola},
  \citenamefont {McKenna}, \citenamefont {Gruenke}, \citenamefont {Patel},
  \citenamefont {Jiang}, \citenamefont {Sarabalis},\ and\ \citenamefont
  {Safavi-Naeini}}]{wollack2021loss}%
  \BibitemOpen
  \bibfield  {author} {\bibinfo {author} {\bibfnamefont {E.~A.}\ \bibnamefont
  {Wollack}}, \bibinfo {author} {\bibfnamefont {A.~Y.}\ \bibnamefont
  {Cleland}}, \bibinfo {author} {\bibfnamefont {P.}~\bibnamefont
  {Arrangoiz-Arriola}}, \bibinfo {author} {\bibfnamefont {T.~P.}\ \bibnamefont
  {McKenna}}, \bibinfo {author} {\bibfnamefont {R.~G.}\ \bibnamefont
  {Gruenke}}, \bibinfo {author} {\bibfnamefont {R.~N.}\ \bibnamefont {Patel}},
  \bibinfo {author} {\bibfnamefont {W.}~\bibnamefont {Jiang}}, \bibinfo
  {author} {\bibfnamefont {C.~J.}\ \bibnamefont {Sarabalis}}, \ and\ \bibinfo
  {author} {\bibfnamefont {A.~H.}\ \bibnamefont {Safavi-Naeini}},\ }\href@noop
  {} {\bibfield  {journal} {\bibinfo  {journal} {Applied Physics Letters}\
  }\textbf {\bibinfo {volume} {118}},\ \bibinfo {pages} {123501} (\bibinfo
  {year} {2021})}\BibitemShut {NoStop}%
\bibitem [{\citenamefont {Li}\ \emph {et~al.}(2021)\citenamefont {Li},
  \citenamefont {Zhang}, \citenamefont {Yang}, \citenamefont {Li},
  \citenamefont {Wang}, \citenamefont {Su}, \citenamefont {Chen}, \citenamefont
  {Li}, \citenamefont {Li}, \citenamefont {Mi} \emph {et~al.}}]{li2021vacuum}%
  \BibitemOpen
  \bibfield  {author} {\bibinfo {author} {\bibfnamefont {X.}~\bibnamefont
  {Li}}, \bibinfo {author} {\bibfnamefont {Y.}~\bibnamefont {Zhang}}, \bibinfo
  {author} {\bibfnamefont {C.}~\bibnamefont {Yang}}, \bibinfo {author}
  {\bibfnamefont {Z.}~\bibnamefont {Li}}, \bibinfo {author} {\bibfnamefont
  {J.}~\bibnamefont {Wang}}, \bibinfo {author} {\bibfnamefont {T.}~\bibnamefont
  {Su}}, \bibinfo {author} {\bibfnamefont {M.}~\bibnamefont {Chen}}, \bibinfo
  {author} {\bibfnamefont {Y.}~\bibnamefont {Li}}, \bibinfo {author}
  {\bibfnamefont {C.}~\bibnamefont {Li}}, \bibinfo {author} {\bibfnamefont
  {Z.}~\bibnamefont {Mi}},  \emph {et~al.},\ }\href@noop {} {\bibfield
  {journal} {\bibinfo  {journal} {Applied Physics Letters}\ }\textbf {\bibinfo
  {volume} {119}},\ \bibinfo {pages} {184003} (\bibinfo {year}
  {2021})}\BibitemShut {NoStop}%
\bibitem [{\citenamefont {Lei}\ \emph {et~al.}(2020)\citenamefont {Lei},
  \citenamefont {Krayzman}, \citenamefont {Ganjam}, \citenamefont {Frunzio},\
  and\ \citenamefont {Schoelkopf}}]{lei2020high}%
  \BibitemOpen
  \bibfield  {author} {\bibinfo {author} {\bibfnamefont {C.~U.}\ \bibnamefont
  {Lei}}, \bibinfo {author} {\bibfnamefont {L.}~\bibnamefont {Krayzman}},
  \bibinfo {author} {\bibfnamefont {S.}~\bibnamefont {Ganjam}}, \bibinfo
  {author} {\bibfnamefont {L.}~\bibnamefont {Frunzio}}, \ and\ \bibinfo
  {author} {\bibfnamefont {R.~J.}\ \bibnamefont {Schoelkopf}},\ }\href@noop {}
  {\bibfield  {journal} {\bibinfo  {journal} {Applied Physics Letters}\
  }\textbf {\bibinfo {volume} {116}},\ \bibinfo {pages} {154002} (\bibinfo
  {year} {2020})}\BibitemShut {NoStop}%
\bibitem [{\citenamefont {Foxen}\ \emph {et~al.}(2017)\citenamefont {Foxen},
  \citenamefont {Mutus}, \citenamefont {Lucero}, \citenamefont {Graff},
  \citenamefont {Megrant}, \citenamefont {Chen}, \citenamefont {Quintana},
  \citenamefont {Burkett}, \citenamefont {Kelly}, \citenamefont {Jeffrey} \emph
  {et~al.}}]{foxen2017qubit}%
  \BibitemOpen
  \bibfield  {author} {\bibinfo {author} {\bibfnamefont {B.}~\bibnamefont
  {Foxen}}, \bibinfo {author} {\bibfnamefont {J.}~\bibnamefont {Mutus}},
  \bibinfo {author} {\bibfnamefont {E.}~\bibnamefont {Lucero}}, \bibinfo
  {author} {\bibfnamefont {R.}~\bibnamefont {Graff}}, \bibinfo {author}
  {\bibfnamefont {A.}~\bibnamefont {Megrant}}, \bibinfo {author} {\bibfnamefont
  {Y.}~\bibnamefont {Chen}}, \bibinfo {author} {\bibfnamefont {C.}~\bibnamefont
  {Quintana}}, \bibinfo {author} {\bibfnamefont {B.}~\bibnamefont {Burkett}},
  \bibinfo {author} {\bibfnamefont {J.}~\bibnamefont {Kelly}}, \bibinfo
  {author} {\bibfnamefont {E.}~\bibnamefont {Jeffrey}},  \emph {et~al.},\
  }\href@noop {} {\bibfield  {journal} {\bibinfo  {journal} {Quantum Science
  and Technology}\ }\textbf {\bibinfo {volume} {3}},\ \bibinfo {pages} {014005}
  (\bibinfo {year} {2017})}\BibitemShut {NoStop}%
\bibitem [{\citenamefont {Rosenberg}\ \emph {et~al.}(2017)\citenamefont
  {Rosenberg}, \citenamefont {Kim}, \citenamefont {Das}, \citenamefont {Yost},
  \citenamefont {Gustavsson}, \citenamefont {Hover}, \citenamefont {Krantz},
  \citenamefont {Melville}, \citenamefont {Racz}, \citenamefont {Samach} \emph
  {et~al.}}]{rosenberg20173d}%
  \BibitemOpen
  \bibfield  {author} {\bibinfo {author} {\bibfnamefont {D.}~\bibnamefont
  {Rosenberg}}, \bibinfo {author} {\bibfnamefont {D.}~\bibnamefont {Kim}},
  \bibinfo {author} {\bibfnamefont {R.}~\bibnamefont {Das}}, \bibinfo {author}
  {\bibfnamefont {D.}~\bibnamefont {Yost}}, \bibinfo {author} {\bibfnamefont
  {S.}~\bibnamefont {Gustavsson}}, \bibinfo {author} {\bibfnamefont
  {D.}~\bibnamefont {Hover}}, \bibinfo {author} {\bibfnamefont
  {P.}~\bibnamefont {Krantz}}, \bibinfo {author} {\bibfnamefont
  {A.}~\bibnamefont {Melville}}, \bibinfo {author} {\bibfnamefont
  {L.}~\bibnamefont {Racz}}, \bibinfo {author} {\bibfnamefont {G.}~\bibnamefont
  {Samach}},  \emph {et~al.},\ }\href@noop {} {\bibfield  {journal} {\bibinfo
  {journal} {npj quantum information}\ }\textbf {\bibinfo {volume} {3}},\
  \bibinfo {pages} {1} (\bibinfo {year} {2017})}\BibitemShut {NoStop}%
\bibitem [{\citenamefont {Kosen}\ \emph {et~al.}(2022)\citenamefont {Kosen},
  \citenamefont {Li}, \citenamefont {Rommel}, \citenamefont {Shiri},
  \citenamefont {Warren}, \citenamefont {Gr{\"o}nberg}, \citenamefont
  {Salonen}, \citenamefont {Abad}, \citenamefont {Bizn{\'a}rov{\'a}},
  \citenamefont {Caputo} \emph {et~al.}}]{kosen2022building}%
  \BibitemOpen
  \bibfield  {author} {\bibinfo {author} {\bibfnamefont {S.}~\bibnamefont
  {Kosen}}, \bibinfo {author} {\bibfnamefont {H.-X.}\ \bibnamefont {Li}},
  \bibinfo {author} {\bibfnamefont {M.}~\bibnamefont {Rommel}}, \bibinfo
  {author} {\bibfnamefont {D.}~\bibnamefont {Shiri}}, \bibinfo {author}
  {\bibfnamefont {C.}~\bibnamefont {Warren}}, \bibinfo {author} {\bibfnamefont
  {L.}~\bibnamefont {Gr{\"o}nberg}}, \bibinfo {author} {\bibfnamefont
  {J.}~\bibnamefont {Salonen}}, \bibinfo {author} {\bibfnamefont
  {T.}~\bibnamefont {Abad}}, \bibinfo {author} {\bibfnamefont {J.}~\bibnamefont
  {Bizn{\'a}rov{\'a}}}, \bibinfo {author} {\bibfnamefont {M.}~\bibnamefont
  {Caputo}},  \emph {et~al.},\ }\href@noop {} {\bibfield  {journal} {\bibinfo
  {journal} {Quantum Science and Technology}\ }\textbf {\bibinfo {volume}
  {7}},\ \bibinfo {pages} {035018} (\bibinfo {year} {2022})}\BibitemShut
  {NoStop}%
\bibitem [{\citenamefont {Harman}(2010)}]{harman2010wire}%
  \BibitemOpen
  \bibfield  {author} {\bibinfo {author} {\bibfnamefont {G.}~\bibnamefont
  {Harman}},\ }\href@noop {} {\emph {\bibinfo {title} {Wire bonding in
  microelectronics}}}\ (\bibinfo  {publisher} {McGraw-Hill Education},\
  \bibinfo {year} {2010})\BibitemShut {NoStop}%
\bibitem [{\citenamefont {Onuki}\ \emph {et~al.}(1987)\citenamefont {Onuki},
  \citenamefont {Suwa}, \citenamefont {Koizumi},\ and\ \citenamefont
  {Iizuka}}]{onuki1987investigation}%
  \BibitemOpen
  \bibfield  {author} {\bibinfo {author} {\bibfnamefont {J.}~\bibnamefont
  {Onuki}}, \bibinfo {author} {\bibfnamefont {M.}~\bibnamefont {Suwa}},
  \bibinfo {author} {\bibfnamefont {M.}~\bibnamefont {Koizumi}}, \ and\
  \bibinfo {author} {\bibfnamefont {T.}~\bibnamefont {Iizuka}},\ }\href@noop {}
  {\bibfield  {journal} {\bibinfo  {journal} {IEEE transactions on components,
  hybrids, and manufacturing technology}\ }\textbf {\bibinfo {volume} {10}},\
  \bibinfo {pages} {242} (\bibinfo {year} {1987})}\BibitemShut {NoStop}%
\bibitem [{\citenamefont {Hizukuri}\ \emph {et~al.}(2001)\citenamefont
  {Hizukuri}, \citenamefont {Watanabe},\ and\ \citenamefont
  {Asano}}]{hizukuri2001dynamic}%
  \BibitemOpen
  \bibfield  {author} {\bibinfo {author} {\bibfnamefont {M.~H.~M.}\
  \bibnamefont {Hizukuri}}, \bibinfo {author} {\bibfnamefont {N.~W.~N.}\
  \bibnamefont {Watanabe}}, \ and\ \bibinfo {author} {\bibfnamefont {T.~A.~T.}\
  \bibnamefont {Asano}},\ }\href@noop {} {\bibfield  {journal} {\bibinfo
  {journal} {Japanese Journal of Applied Physics}\ }\textbf {\bibinfo {volume}
  {40}},\ \bibinfo {pages} {3044} (\bibinfo {year} {2001})}\BibitemShut
  {NoStop}%
\bibitem [{\citenamefont {Tani}\ \emph {et~al.}(2010)\citenamefont {Tani},
  \citenamefont {Watanabe}, \citenamefont {Nishimura}, \citenamefont {Kachi},
  \citenamefont {Katada},\ and\ \citenamefont {Sugiura}}]{tani2010development}%
  \BibitemOpen
  \bibfield  {author} {\bibinfo {author} {\bibfnamefont {M.}~\bibnamefont
  {Tani}}, \bibinfo {author} {\bibfnamefont {H.}~\bibnamefont {Watanabe}},
  \bibinfo {author} {\bibfnamefont {A.}~\bibnamefont {Nishimura}}, \bibinfo
  {author} {\bibfnamefont {S.}~\bibnamefont {Kachi}}, \bibinfo {author}
  {\bibfnamefont {N.}~\bibnamefont {Katada}}, \ and\ \bibinfo {author}
  {\bibfnamefont {S.}~\bibnamefont {Sugiura}},\ }\href@noop {} {\bibfield
  {journal} {\bibinfo  {journal} {Fujitsu Ten Tech. J.}\ }\textbf {\bibinfo
  {volume} {34}},\ \bibinfo {pages} {19} (\bibinfo {year} {2010})}\BibitemShut
  {NoStop}%
\bibitem [{\citenamefont {Shuto}\ and\ \citenamefont
  {Asano}(2015)}]{shuto2015situ}%
  \BibitemOpen
  \bibfield  {author} {\bibinfo {author} {\bibfnamefont {T.}~\bibnamefont
  {Shuto}}\ and\ \bibinfo {author} {\bibfnamefont {T.}~\bibnamefont {Asano}},\
  }\href@noop {} {\bibfield  {journal} {\bibinfo  {journal} {Japanese Journal
  of Applied Physics}\ }\textbf {\bibinfo {volume} {54}},\ \bibinfo {pages}
  {030204} (\bibinfo {year} {2015})}\BibitemShut {NoStop}%
\bibitem [{\citenamefont {Wenner}\ \emph {et~al.}(2011)\citenamefont {Wenner},
  \citenamefont {Neeley}, \citenamefont {Bialczak}, \citenamefont {Lenander},
  \citenamefont {Lucero}, \citenamefont {O'connell}, \citenamefont {Sank},
  \citenamefont {Wang}, \citenamefont {Weides}, \citenamefont {Cleland},\ and\
  \citenamefont {Martinis}}]{Wenner2011}%
  \BibitemOpen
  \bibfield  {author} {\bibinfo {author} {\bibfnamefont {J.}~\bibnamefont
  {Wenner}}, \bibinfo {author} {\bibfnamefont {M.}~\bibnamefont {Neeley}},
  \bibinfo {author} {\bibfnamefont {R.~C.}\ \bibnamefont {Bialczak}}, \bibinfo
  {author} {\bibfnamefont {M.}~\bibnamefont {Lenander}}, \bibinfo {author}
  {\bibfnamefont {E.}~\bibnamefont {Lucero}}, \bibinfo {author} {\bibfnamefont
  {A.~D.}\ \bibnamefont {O'connell}}, \bibinfo {author} {\bibfnamefont
  {D.}~\bibnamefont {Sank}}, \bibinfo {author} {\bibfnamefont {H.}~\bibnamefont
  {Wang}}, \bibinfo {author} {\bibfnamefont {M.}~\bibnamefont {Weides}},
  \bibinfo {author} {\bibfnamefont {A.~N.}\ \bibnamefont {Cleland}}, \ and\
  \bibinfo {author} {\bibfnamefont {J.~M.}\ \bibnamefont {Martinis}},\
  }\href@noop {} {\bibfield  {journal} {\bibinfo  {journal} {Supercond. Sci.
  Technol}\ }\textbf {\bibinfo {volume} {24}},\ \bibinfo {pages} {65001}
  (\bibinfo {year} {2011})}\BibitemShut {NoStop}%
\bibitem [{\citenamefont {Huang}\ \emph {et~al.}(2021)\citenamefont {Huang},
  \citenamefont {Lienhard}, \citenamefont {Calusine}, \citenamefont
  {Veps{\"a}l{\"a}inen}, \citenamefont {Braum{\"u}ller}, \citenamefont {Kim},
  \citenamefont {Melville}, \citenamefont {Niedzielski}, \citenamefont {Yoder},
  \citenamefont {Kannan} \emph {et~al.}}]{huang2021microwave}%
  \BibitemOpen
  \bibfield  {author} {\bibinfo {author} {\bibfnamefont {S.}~\bibnamefont
  {Huang}}, \bibinfo {author} {\bibfnamefont {B.}~\bibnamefont {Lienhard}},
  \bibinfo {author} {\bibfnamefont {G.}~\bibnamefont {Calusine}}, \bibinfo
  {author} {\bibfnamefont {A.}~\bibnamefont {Veps{\"a}l{\"a}inen}}, \bibinfo
  {author} {\bibfnamefont {J.}~\bibnamefont {Braum{\"u}ller}}, \bibinfo
  {author} {\bibfnamefont {D.~K.}\ \bibnamefont {Kim}}, \bibinfo {author}
  {\bibfnamefont {A.~J.}\ \bibnamefont {Melville}}, \bibinfo {author}
  {\bibfnamefont {B.~M.}\ \bibnamefont {Niedzielski}}, \bibinfo {author}
  {\bibfnamefont {J.~L.}\ \bibnamefont {Yoder}}, \bibinfo {author}
  {\bibfnamefont {B.}~\bibnamefont {Kannan}},  \emph {et~al.},\ }\href@noop {}
  {\bibfield  {journal} {\bibinfo  {journal} {PRX Quantum}\ }\textbf {\bibinfo
  {volume} {2}},\ \bibinfo {pages} {020306} (\bibinfo {year}
  {2021})}\BibitemShut {NoStop}%
\bibitem [{\citenamefont {Zhong}\ \emph {et~al.}(2021)\citenamefont {Zhong},
  \citenamefont {Chang}, \citenamefont {Bienfait}, \citenamefont {Dumur},
  \citenamefont {Chou}, \citenamefont {Conner}, \citenamefont {Grebel},
  \citenamefont {Povey}, \citenamefont {Yan}, \citenamefont {Schuster} \emph
  {et~al.}}]{zhong2021deterministic}%
  \BibitemOpen
  \bibfield  {author} {\bibinfo {author} {\bibfnamefont {Y.}~\bibnamefont
  {Zhong}}, \bibinfo {author} {\bibfnamefont {H.-S.}\ \bibnamefont {Chang}},
  \bibinfo {author} {\bibfnamefont {A.}~\bibnamefont {Bienfait}}, \bibinfo
  {author} {\bibfnamefont {{\'E}.}~\bibnamefont {Dumur}}, \bibinfo {author}
  {\bibfnamefont {M.-H.}\ \bibnamefont {Chou}}, \bibinfo {author}
  {\bibfnamefont {C.~R.}\ \bibnamefont {Conner}}, \bibinfo {author}
  {\bibfnamefont {J.}~\bibnamefont {Grebel}}, \bibinfo {author} {\bibfnamefont
  {R.~G.}\ \bibnamefont {Povey}}, \bibinfo {author} {\bibfnamefont
  {H.}~\bibnamefont {Yan}}, \bibinfo {author} {\bibfnamefont {D.~I.}\
  \bibnamefont {Schuster}},  \emph {et~al.},\ }\href@noop {} {\bibfield
  {journal} {\bibinfo  {journal} {Nature}\ }\textbf {\bibinfo {volume} {590}},\
  \bibinfo {pages} {571} (\bibinfo {year} {2021})}\BibitemShut {NoStop}%
\end{thebibliography}%

\section*{Supplemental Information}

\renewcommand{\figurename}{FIG.}
\renewcommand{\thefigure}{S\arabic{figure}}
\setcounter{figure}{0}

\renewcommand{\tablename}{TABLE.}
\renewcommand{\thetable}{S\arabic{table}}
\setcounter{table}{0}

\subsection*{System Hamiltonian and derivation of the cavity scattering parameters}
The actual system consists of multiple frequency-multiplexed mechanical modes, a microwave mode, and a cavity mode. The Hamiltonian describing the system is 
\begin{eqnarray} \label{eqS1}
    \hat H / \hbar&=&\omega_\text{c} \hat{c}^{\dagger} \hat{c}+\omega_\text{a} \hat{a}^{\dagger} \hat{a}+ \sum_{n=1}^{N}\omega_{\text{m},n} \hat{b}^{\dagger}_{n} \hat{b}_{n} \notag\\
    &&+ \sum_{n=1}^{N} g_{\text{ab},n}\left(\hat{a}^{\dagger} \hat{b}_{n}+\hat{b}_{n}^{\dagger} \hat{a}\right)+g_\text{ac}\left(\hat{a}^{\dagger} \hat{c}+\hat{c}^{\dagger} \hat{a}\right),
\end{eqnarray}
where $\omega_{c}$, $\omega_{a}$, $\omega_{\text{m},n}$ and $\hat{c}$, $\hat{a}$, $\hat{b}_{n}$ represent the frequencies and annihilation operators of the cavity mode, the microwave mode, and the $N$ mechanical modes, respectively. The coupling rate between the cavity and the microwave modes, and the coupling rate between the microwave and the mechanical modes are represented by $g_\text{ac}$ and $g_{\text{ab},n}$, respectively.

We can solve these equations classically, assuming a coherent state input field into the $c$ mode, to derive a set of coupled differential equations of motion. These linear differential equations, in the Fourier domain are:
\begin{eqnarray} \label{eqS2}
-i \omega b_n(\omega) &=&  -\left( i \omega_{\text{b},n} + \frac{\gamma_{\text{b},n}}{2} \right) b_{n}(\omega)  - i g_{\text{ab},n} a(\omega)
\label{eqS3}\\
-i \omega a(\omega) &=& -\left( i \omega_\text{a} + \frac{\kappa_{\text{a,i}}}{2} \right) a(\omega)  -\nonumber\\&&~~ i \sum_{n=1}^{N} g_{\text{ab},n} b_{n}(\omega) - i g_\text{ac} c(\omega)\\
\label{eqS4}
-i \omega c(\omega) &=& -\left( i \omega_\text{c} + \frac{\kappa_\text{c}}{2} \right) c(\omega)- \nonumber\\&&~~i g_\text{ac} a(\omega)  - \sqrt{\kappa_\text{c,1}} c_\text{in,1}(\omega) ,
\end{eqnarray}
where $\kappa_\text{c} = (\kappa_\text{c,1} + \kappa_\text{c,2} + \kappa_\text{c,i})$. The input-output boundary conditions for the two ports of the cavity are given by
\begin{eqnarray} \label{eqS5}
c_\text{out,1}(\omega) &=& c_\text{in,1}(\omega) - \sqrt{\kappa_\text{c,1}} c(\omega)\\\label{eqS6}
c_\text{out,2}(\omega) &=&  - \sqrt{\kappa_\text{c,2}} c(\omega).
\end{eqnarray}

The above equations (Eqs. \ref{eqS2}-\ref{eqS6}) can be solved for the reflection $r(\omega)$ and transmission $t(\omega)$ coefficients. After some algebra, we find that
\begin{eqnarray} \label{eqS7}
r(\omega) &\equiv& \frac{c_\text{out,1}(\omega)}{c_\text{in,1}(\omega)}  \notag\\
&=& 1 - \frac{\kappa_\text{c,1}}{-i \Delta_\text{c} + \frac{\kappa_\text{c}}{2} + \frac{g_\text{ac}^2}{-i \Delta_\text{a} + \frac{\kappa_{\text{a,i}}}{2} + \sum_{n=1}^{N} \left(\frac{g_{\text{ab},n}^2}{-i \Delta_{\text{b},n} + \frac{\gamma_{\text{b},n}}{2}}\right) }}
\end{eqnarray}
and
\begin{eqnarray} \label{eqS8}
t(\omega) &\equiv& \frac{c_\text{out,2}(\omega)}{c_\text{in,1}(\omega)}  \notag\\
&=& \frac{-\sqrt{\kappa_\text{c,1}}\sqrt{\kappa_\text{c,2}}}{-i \Delta_\text{c} + \frac{\kappa_\text{c}}{2} + \frac{g_\text{ac}^2}{-i \Delta_\text{a} + \frac{\kappa_{\text{a,i}}}{2} + \sum_{n=1}^{N} \left(\frac{g_{\text{ab},n}^2}{-i \Delta_{\text{b},n} + \frac{\gamma_{\text{b},n}}{2}}\right) }}.
\end{eqnarray}
where $\Delta_{\text{b},n} \equiv \omega - \omega_{\text{b},n}$, $\Delta_\text{a} \equiv \omega - \omega_\text{a}$, $\Delta_\text{c} \equiv \omega - \omega_\text{c}$.

\subsection*{Fitting the data}
Because the attenuation and amplification chains in the dilution refrigerator are not calibrated accurately, we shift the theoretic transmission coefficient in Eq. \ref{eqS8} with an artificial free offset parameter to take into account the attenuation and gain in the measurement chain
\begin{eqnarray} \label{eqS9}
S_{21}(\omega) &=&  C_\text{offset} \times t(\omega).
\end{eqnarray}

\begin{figure*}[tb]
\includegraphics[scale=1]{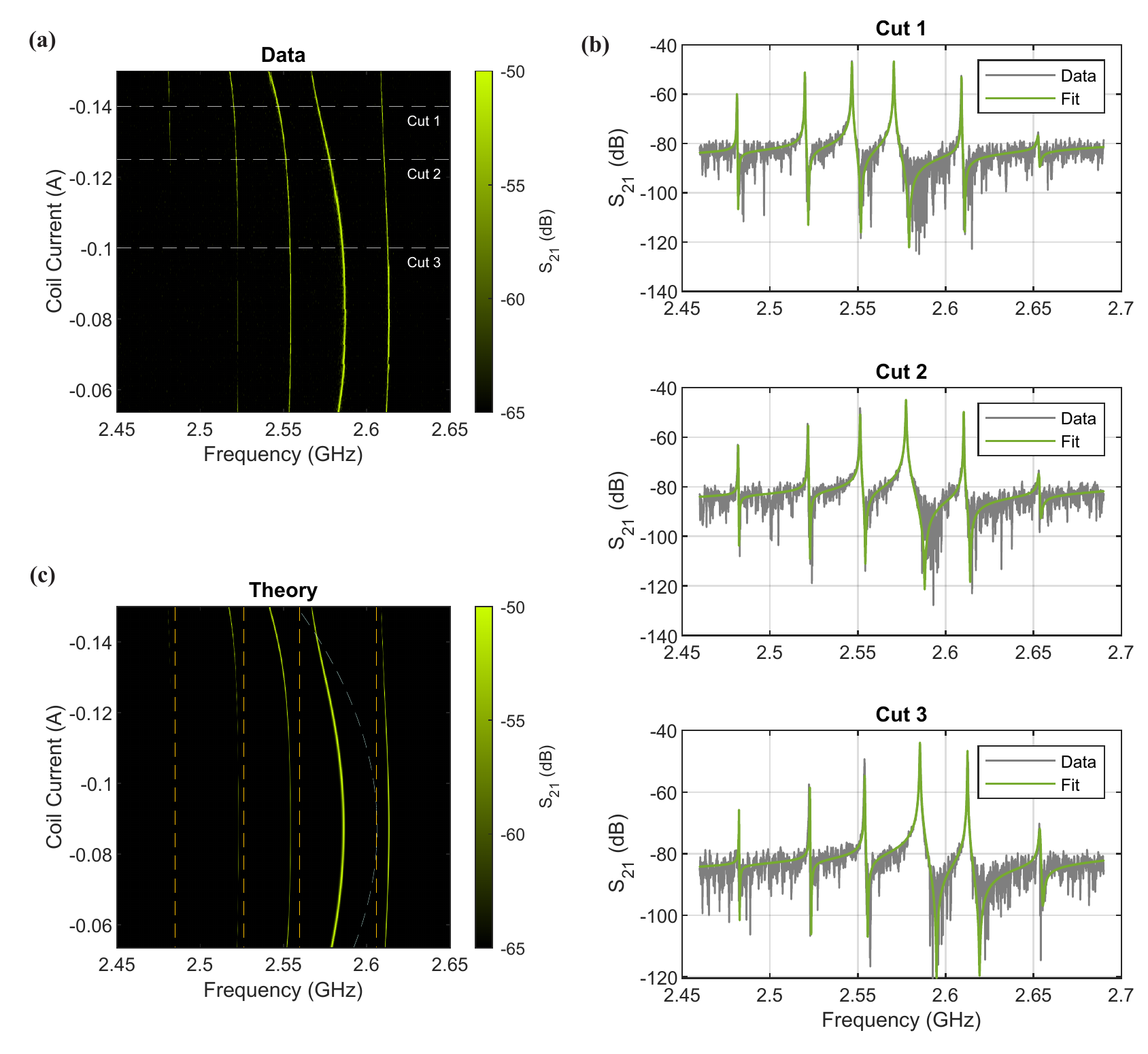}
\caption{\label{SI-fig:fcbbfit} \textbf{Extraction of theory fit parameters for a flip-chip bump bonded chip pair.} \textbf{(a)} Transmission measurement as a function of the applied magnetic field represented by the coil current. Three cuts used for the fitting are shown in white dashed lines. \textbf{(b)} Transmission measurement data (grey) overlayed with the fitted coupled mode input-output theory (green) at the three cuts. Fitting is done using a genetic algorithm.  \textbf{(c)} Reconstructed theory plot with the individual mechanical modes shown as yellow dashed lines and the microwave mode shown as a blue dashed line.}
\end{figure*}

Because the cavity mode is placed relatively far from the microwave and mechanical modes, we use the wide-scan data to fit the cavity mode and extract the cavity parameters independently. We then use the scattering parameter in Eq. \ref{eqS9} to fit the linear $S_{21}$ data by minimizing the cost function over the system parameters $\{x_\text{mech + mw}\}$ using a generic algorithm at three different cuts on the coil sweep plot (see Fig.~\ref{SI-fig:fcbbfit}(a) for the flip-chip bump bond case, Fig.~\ref{SI-fig:xcwbfit}(a) for the cross-chip wirebond case),

\begin{eqnarray*}
f(\{x_\text{sys}\}) &=&  \sum_{cuts} \sum_{j} [ S_{21}(\{x_\text{mech + mw}\}, \omega_{j}) - S_{21, \text{data}}(\omega_{j}) ]^2
\end{eqnarray*}
with the mechanical modes' parameters ($\omega_{\text{b},n}$, $\gamma_{\text{b},n}$) kept fixed between the three cuts. The loss rate $\kappa_{\text{a,i}}$ of the microwave mode is allowed to vary between the three cuts. The fit results for the flip-chip bump bond and the cross-chip wirebond cases are shown in Table.~\ref{tab:fcbbparameters} and Table.~\ref{tab:xcwbparameters}, respectively. We can use these fit results to extract the microwave mode tuning and reconstruct the full coil sweep as shown in Fig.~\ref{SI-fig:fcbbfit}(c) and Fig.~\ref{SI-fig:xcwbfit}(c).

\subsection*{Cross-chip wirebond}
\begin{figure*}[tb]
\includegraphics[scale=0.9]{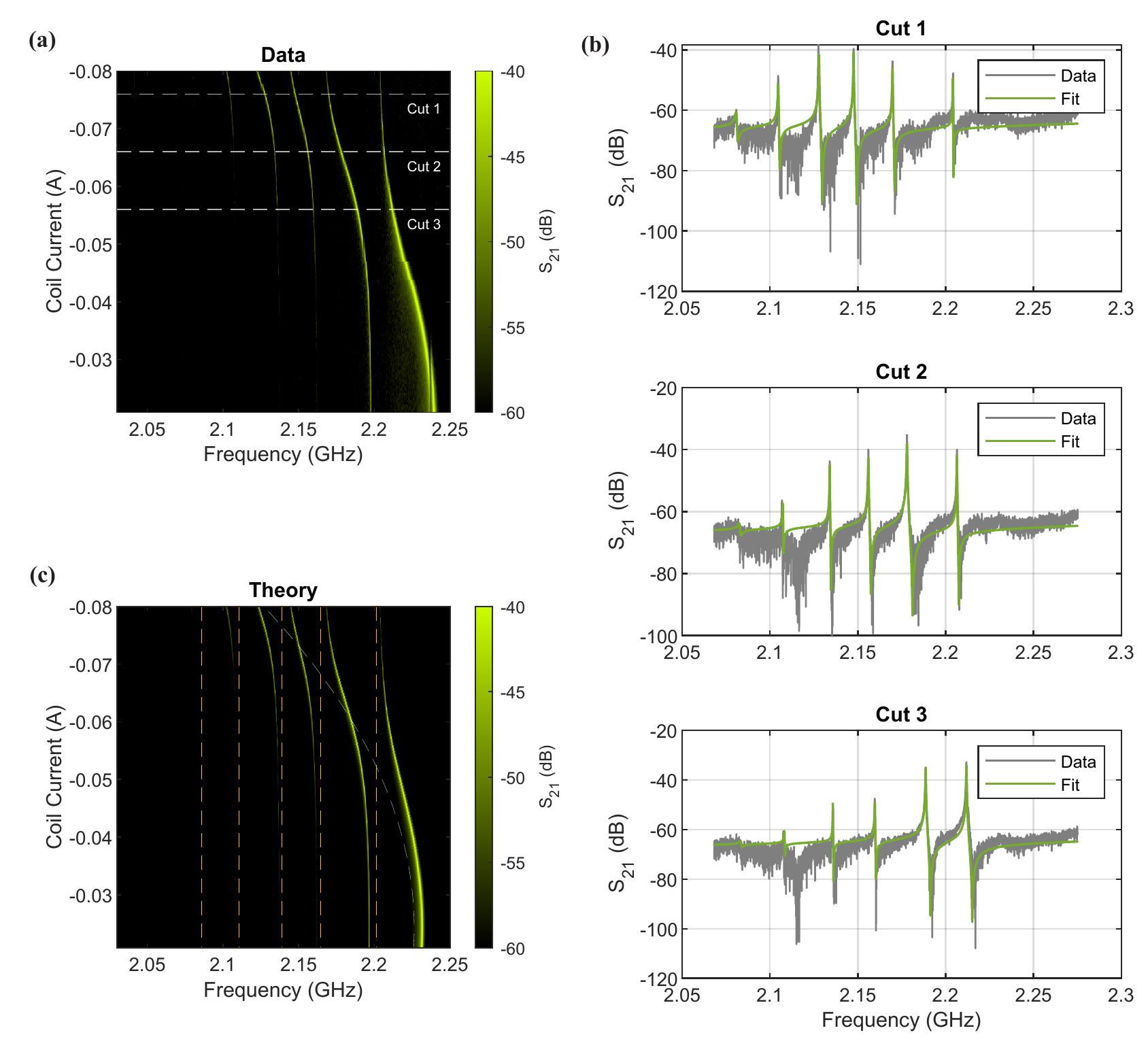}
\caption{\label{SI-fig:xcwbfit} \textbf{Extraction of theory fit parameters for cross-chip wirebonded chip pair.} \textbf{(a)} Transmission measurement as a function of the applied magnetic field represented by the coil current. Three cuts used for the fitting are shown in white dashed lines. \textbf{(b)} Transmission measurement data (grey) overlayed with the fitted coupled mode input-output theory (green) at the three cuts. Fitting is done using a genetic algorithm.  \textbf{(c)} Reconstructed theory plot with the individual mechanical modes shown as yellow dashed lines and the microwave mode shown as a blue dashed line.}
\end{figure*}

In addition to the flip-chip integration approach discussed in the paper, we pursued an alternative crude integration approach for comparison. In this approach, the microwave and the mechanics chips are placed adjacent to each other in one plane and a cross-chip wirebond connects the two. The cryogenic measurement for this cross-chip wirebonded chip pair is shown in Fig.~\ref{SI-fig:xcwbfit}(a). By following a similar fitting approach as discussed in the prior section, we extract microwave-mechanics coupling rates of $\sim \SI{12}-\SI{14}{\mega\hertz}$ (Fig.~\ref{SI-fig:xcwbfit}, Table.~\ref{tab:xcwbparameters}). 

\begin{figure*}[tb]
\includegraphics[scale=1]{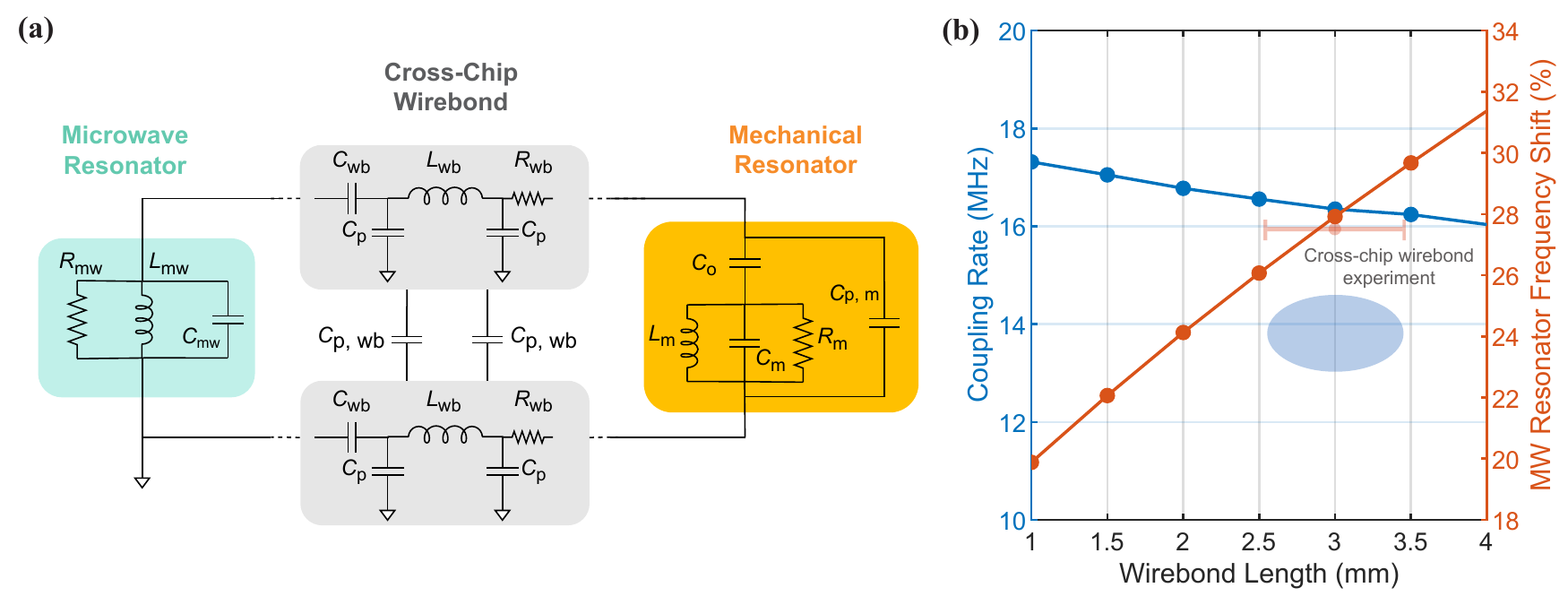}
\caption{\label{SI-fig:xcwbmodel} \textbf{Cross-chip wirebond model.} \textbf{(a)} Lumped element circuit model of the cross-chip wirebonds between the mechanical mode and the microwave mode. The mechanical mode is modeled as a $RLC$ circuit (with parameters $R_{\text m}$, $L_{\text{m}}$, $C_{\text m}$, $C_{\text{o}}$, and $C_{\text{p,m}}$). The microwave mode is also modeled as a $RLC$ circuit (with parameters $R_{\text m}$, $L_{\text{m}}$, $C_{\text m}$, and a characteristic impedance of $Z \sim \SI{400}{\ohm}$). The signal and ground cross-chip wirebonds are modeled as series inductance $ L_{\text{wb}}$ with parasitic capacitance to the ground plane $C_{\text p}$ in parallel, a contact capacitance $C_{\text{wb}}$ and a contact resistance $R_{\text{wb}}$ in series. The signal and ground wirebonds have a mutual capacitance $C_{\text{p,wb}}$ between them. \textbf{(b)} Coupling rate between the mechanical and microwave modes, and the microwave mode frequency shift due to the wirebonds as a function of wirebond length. The shaded region and the horizontal bar show the estimated coupling rate and microwave frequency shift in this cross-chip wirebond pair experiment. While the coupling rate is not very sensitive to the wirebond length, the shift in microwave mode frequency is very sensitive to the wirebond length.}
\end{figure*}

To understand the impact of the wirebond on the coupling rate and the microwave mode frequency shift, we consider a lumped element circuit model of the wirebonds between the mechanical mode and the microwave mode, as shown in Fig.~\ref{SI-fig:xcwbmodel}(a). The mechanical mode is modeled as a $RLC$ circuit with parameters $L_{\text{m}} = \SI{2.73}{\nano\henry}$, $C_{\text{m}} = \SI{1.83}{\femto\farad} $, $R_{\text{m}} = \SI{884}{\mega\ohm}$, $C_{\text{o}} = \SI{337}{\atto\farad}$, and $C_{\text{p,m}} = \SI{50}{\femto\farad}$, estimated from a combination of finite-element simulation and room temperature reflection measurements. The microwave mode is also modeled as a $RLC$ circuit with parameters $L_{\text{mw}} = \SI{20}{\nano\henry}$, $C_{\text{mw}} = \SI{130.8}{\femto\farad}$, and $R_{\text{mw}} = \SI{37}{\mega\ohm}$, estimated from a combination of finite-element simulation of the capacitances and cryogenic measurement of the microwave mode frequency. We model the signal and ground cross-chip wirebonds as series inductance $L_{\text{wb}}$ with parasitic capacitance to the chip plane $C_{\text p}$ in parallel, and a contact capacitance $C_{\text{wb}}$ and a contact resistance $R_{\text{wb}}$ in series. The wirebonds have a mutual capacitance of $C_{\text{p,wb}}$ between them. For a $\SI{25}{\micro\meter}$ diameter wire, the inductance roughly scales as $L_{\text{wb}}$ $\sim\SI{1}{\nano\henry}/\SI{}{\milli\meter}$ \cite{Wenner2011, huang2021microwave}. The remaining wirebond parameters are estimated using finite-element electrostatic analysis with nominal values of $C_{\text{p}}$ $\sim\SI{3.9}{\femto\farad}$, $C_{\text{wb}}$ $\sim\SI{20}{\pico\farad}$, $R_{\text{wb}}$ $\sim\SI{0.4}{\ohm}$~\cite{zhong2021deterministic}, and  $C_{\text{p,wb}}$ $\sim\SI{7}{\femto\farad}/\SI{}{\milli\meter}$.

The lumped element model shows a small decrease in the microwave-mechanics coupling rate as a function of the wirebond length as shown in Fig.~\ref{SI-fig:xcwbmodel}(b). This is expected because the capacitance of the wirebonds with the ground and to each other scales with the length. More importantly, we learn that the post-integration frequency shift of the microwave mode is very sensitive to the length of the wirebond. Due to the lack of control over the length and shape of a wirebond, the post-integration frequency shift can vary significantly. Because we want the microwave mode to be appropriately parked with respect to the cavity mode and the mechanical modes, the variance in the microwave frequency shift makes it difficult to select, a priori,  a microwave resonator for integration with the mechanics chips.

\subsection*{Experimental setup}
\begin{figure*}
\includegraphics[scale=1]{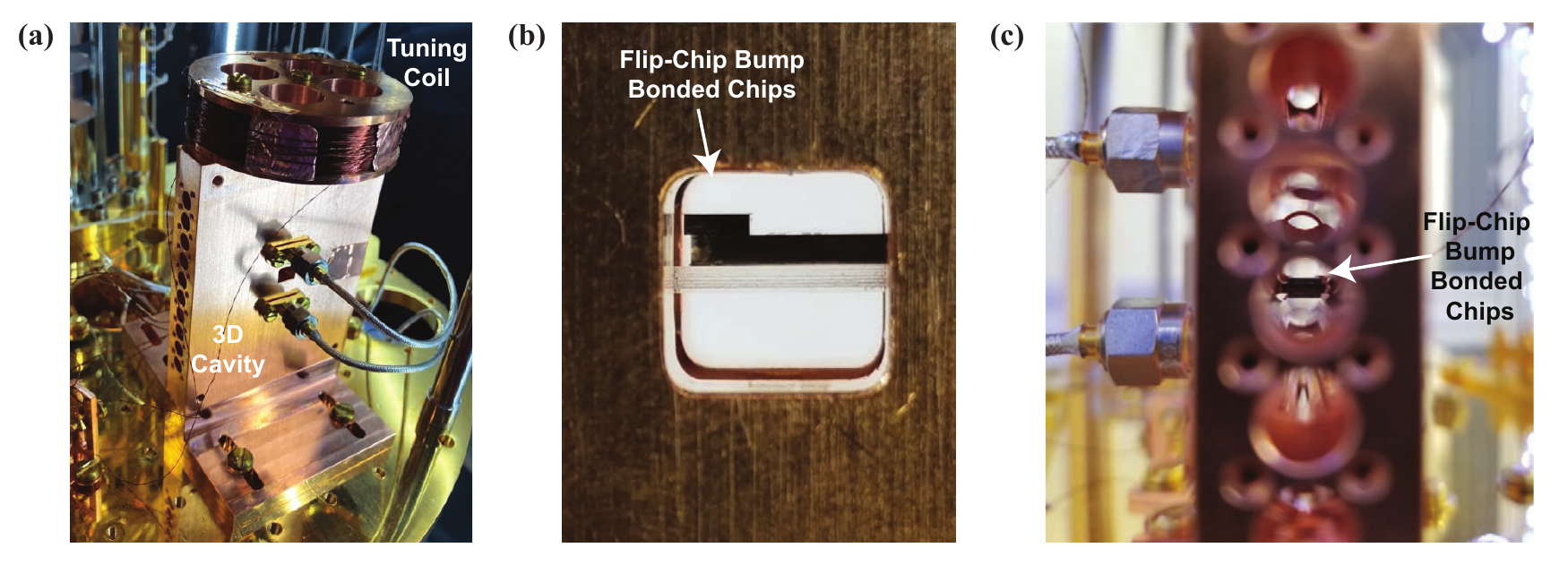}
\caption{\label{SI-fig:setup} \textbf{Experimental setup.} \textbf{(a)} 3D cavity with a homemade tuning coil mounted on top of it. The 3D cavity has two external ports that enable transmission measurement. The tuning coil generates $\sim\SI{0.01}{\milli\tesla\per\milli\ampere}$ at the chip placed inside the 3D cavity. \textbf{(b)} Side view through the optical port on the 3D cavity. The flip-chip bump bonded pairs of microwave-mechanics chips can be seen mounted on a sapphire carrier. \textbf{(c)} View looking in through the drilled hole of the 3D cavity.}
\end{figure*}

Fig.~\ref{SI-fig:setup} shows the experimental setup in this work. The 3D cavity is made from oxygen-free high thermal conductivity copper (OFHC) to allow magnetic field to penetrate into the cavity for kinetic-inductance tuning of the microwave mode. A home-made tuning coil (\SI{50}{\milli\meter} diameter and $\sim 2000$ turns of NbTi wire) is mounted on top of the 3D cavity and generates $\sim\SI{0.01}{\milli\tesla\per\milli\ampere}$ at the chip placed inside the cavity. Due to the superconducting NbTi wire coil, we observe negligible heating in the dilution refrigerator with up to \SI{200}{\milli\ampere} continuous current. 

The 3D cavity has two ports that enable both the reflection and transmission spectrum measurement. Additionally, the cavity has two optical ports (Fig.~\ref{SI-fig:setup}(a)) for coupling the chip to an optical fiber for microwave-to-optical transduction experiments. The optical port size is chosen such that the cutoff frequency of the waveguide mode is much higher than the cavity modes. Looking through the optical port (Fig.~\ref{SI-fig:setup}(b)), we can see the flip-chip microwave-mechanics pair glued on a sapphire carrier. 

The 3D cavity is mounted on the \SI{7}{\milli\kelvin} stage of a Bluefors dilution refrigerator. The microwave input line is attenuated to reduce thermal noise. The microwave signal from the cavity is amplified by two room-temperature low-noise amplifiers. Transmission measurements are performed using a vector network analyzer (VNA, Rohde-Schwarz ZNB20).

\begin{table*}[!htbp]
\begin{center}
\begin{tabular}{c  c }
\hline
Parameter & Value\\
\hline
$\omega_\text{c}/2\pi$ & \SI{2.923}{\giga\hertz}\\
$\kappa_\text{c}/2\pi$ & \SI{444}{\kilo\hertz}\\
$\omega_\text{a}/2\pi$ & \SI{2.572}, \SI{2.589}, \SI{2.604}{\giga\hertz}\\
$\kappa_\text{a,i}/2\pi$ & \SI{295}, \SI{346}, \SI{339}{\kilo\hertz}\\
$\omega_\text{m}/2\pi$ & \SI{2.485}, \SI{2.526}, \SI{2.559}, \SI{2.606}, \SI{2.651}{\giga\hertz}\\
$\gammai/2\pi$ & \SI{81}, \SI{80}, \SI{149}, \SI{72}, \SI{836}{\kilo\hertz}\\
$g_\text{ac}/2\pi$ & \SI{83.466}{\mega\hertz}\\
$g_\text{ab}/2\pi$ & \SI{15.314}, \SI{14.364}, \SI{14.255}, \SI{13.590}, \SI{13.633}{\mega\hertz}\\
\hline
\end{tabular}
\caption{\label{tab:fcbbparameters}\textbf{Extracted theory fit parameters for flip-chip bump bonded chip pair.} The three values for the microwave resonator frequency and linewidth correspond to the three fitting cuts.}
\end{center}
\end{table*}

\begin{table*}[!htbp]
\begin{center}
\begin{tabular}{c  c }
\hline
Parameter & Value\\
\hline
$\omega_\text{c}/2\pi$ & \SI{2.837}{\giga\hertz}\\
$\kappa_\text{c}/2\pi$ & \SI{1026}{\kilo\hertz}\\
$\omega_\text{a}/2\pi$ & \SI{2.141}, \SI{2.171}, \SI{2.194}{\giga\hertz}\\
$\kappa_\text{a,i}/2\pi$ & \SI{346}, \SI{342}, \SI{239}{\kilo\hertz}\\
$\omega_\text{m}/2\pi$ & \SI{2.086}, \SI{2.111}, \SI{2.139}, \SI{2.164}, \SI{2.201}{\giga\hertz}\\
$\gammai/2\pi$ & \SI{1073}, \SI{270}, \SI{88}, \SI{207}, \SI{98}{\kilo\hertz}\\
$g_\text{ac}/2\pi$ & \SI{68.295}{\mega\hertz}\\
$g_\text{ab}/2\pi$ & \SI{14.234}, \SI{13.549}, \SI{12.774}, \SI{13.026}, \SI{12.883}{\mega\hertz}\\
\hline
\end{tabular}
\caption{\label{tab:xcwbparameters}\textbf{Extracted theory fit parameters for cross-chip wirebonded chip pairs.} The three values for the microwave resonator frequency and linewidth correspond to the three fitting cuts.}
\end{center}
\end{table*}

\end{document}


\preprint{APS/123-QED}

\title{Heterogeneous integration of high kinetic inductance resonator with thin-film lithium niobate nanomechanical resonators - Supplementary Information}

\date{\today}

\maketitle


\section*{Supplementary Information}

\subsection*{Cross-Chip Wirebond Results}

\subsection*{Cross-chip Wirebond Model}

\subsection*{Fabrication}

\subsection*{3D Cavity Characterization}

\nocite{*}

\bibliography{ref}